\documentclass[aps,prx,twocolumn,superscriptaddress,longbibliography]{revtex4-2}

\usepackage{graphicx}
\usepackage{dcolumn}
\usepackage{bm}
\usepackage[T1]{fontenc}
\usepackage{textcomp}
\usepackage{amsmath}
\usepackage{mathrsfs}
\usepackage{bm}
\usepackage{physics}
\usepackage{url}
\usepackage{amssymb}
\usepackage{lineno} 
\usepackage{dsfont}
\usepackage{xcolor}
\usepackage{comment}
\PassOptionsToPackage{hyphens}{url}\usepackage[colorlinks,bookmarks=false,citecolor=blue,linkcolor=blue,urlcolor=blue]{hyperref}
\usepackage{braket}
\usepackage{stackrel}
\usepackage{breqn}
\usepackage{caption}
\usepackage{subcaption}
\usepackage{stmaryrd}
\usepackage{wrapfig}
\usepackage{multirow}
\usepackage[normalem]{ulem}
\usepackage[nottoc,numbib]{tocbibind}
\usepackage{cleveref}
\usepackage{siunitx}
\usepackage{xr}
\usepackage{adjustbox}
\usepackage{csquotes}
\usepackage{subfiles}
\usepackage{lineno}

\DeclareMathOperator{\atan2}{atan2}
\DeclareSIUnit\pixel{pixel}

\begin{document}

\title{A sublattice Stokes polarimeter for bipartite photonic lattices}

\author{M. Guillot}

\author{C. Blanchard}

\author{N. Pernet}

\author{M. Morassi}

\author{A. Lema\^itre}

\author{L. Le Gratiet}

\author{A. Harouri}

\author{I. Sagnes}

\author{J. Bloch}

\author{S. Ravets}
\email[]{sylvain.ravets@c2n.upsaclay.fr}

\affiliation{Center for Nanoscience and Nanotechnology$,$ CNRS$,$ Paris--Saclay university$,$ 91120 Palaiseau$,$ France}



\begin{abstract}

\bigskip

\textbf{The concept of pseudo-spin provides a general framework for describing physical systems featuring two-component spinors, including light polarization, sublattice degrees of freedom in bipartite lattices, and valley polarization in 2D materials. In all cases, the pseudo-spin can be mapped to a Stokes vector on the Poincaré sphere. Stokes polarimeters for measuring the polarization of light are a powerful tool with a wide range of applications both in classical and quantum science. Generalizing Stokes polarimetry to other spinor degrees of freedom is thus a challenge of prime importance. Here, we introduce and demonstrate a Stokes polarimeter for the sublattice polarization in a bipartite photonic lattice. Our method relies on $k$-space photoluminescence intensity measurements under controlled phase shifts and attenuations applied independently to each sublattice. We implement our method using honeycomb arrays of coupled microcavities realizing photonic analogs of graphene and hexagonal boron nitride. Using our sublattice polarimeter, we reconstruct the Bloch modes in amplitude and phase across the Brillouin zone, achieving sub-linewidth precision in the determination of their eigenenergies, including near band touching points. This enables full access to the system Bloch Hamiltonian and quantum geometric tensor. Our approach can readily be extended to more complex systems with additional internal degrees of freedom, enabling experimental investigations of trigonal warping, Chern insulating phases, and Euler-class topology in multigap systems.}

\end{abstract}

\maketitle

Electronic band structures of ordered solids, tracing the energy of electrons versus momentum $k$, provide a deep understanding of many optoelectronic properties of condensed matter. Engineering the shape and energy gaps of semiconductor band structures has led to the development of a large class of electronic and optoelectronic devices. Going beyond energy bands, pioneering developments in topological physics have highlighted the importance of the geometry and symmetries of the space spanned by a system eigenstates, in capturing some physical phenomena~\cite{Hasan2010,Haldane2017}. Manifestations of topology include the geometric phase accumulated while navigating through $k$-space during the adiabatic evolution of a Hamiltonian~\cite{Berry1984}, the existence of topological invariants in the bulk of a material~\cite{Thouless1982}, or the presence of edge states at the interface between two materials~\cite{Kane2005}. This physics is encoded in the quantum geometric tensor~\cite{Provost1980}, defined from the system eigenstates. Its imaginary part captures the Berry curvature, which gives rise to anomalous Hall drift~\cite{Nagaosa2010}, spin-Hall~\cite{Sinova2015}, and valley-Hall effects~\cite{Mak2014}. Its real part is the quantum metric, which has been shown to play a key role in influencing superfluidity in flatbands~\cite{Peotta2015,Torma2022}, modifying the electronic orbital magnetic susceptibility~\cite{Piechon2016}, or inducing nonlinear transport~\cite{Wang2023,Gao2023}.

Experimentally mapping the full geometry of Bloch eigenstates remains a major challenge, particularly in systems featuring multiple pseudo-spin degrees of freedom. Recent breakthroughs have demonstrated the feasibility of measuring the quantum metric in real materials, by partially reconstructing the Bloch wavefunctions using angle-resolved photoemission spectroscopy~\cite{Kang2025,Kim2025}. In synthetic systems with a single spinor degree of freedom, impressive achievements have been obtained in measuring Bloch eigenstates and probing band topology, owing to a high degree of control and tunability. Eigenstate tomography for the sublattice degree of freedom was realized in atomic~\cite{Li2016,Tarnowski2017,Flaschner2016}, Floquet~\cite{Wimmer2017,Lechevalier2021} or synthetic frequency dimension ~\cite{Dutt2019,Chenier2024,Cheng2025} lattices, and for the spin degree of freedom in atomic or solid-state two-level systems~\cite{Yi2023,Yu2019,Tan2019}. The polarization of light offers a compelling pseudo-spin degree of freedom that can be conveniently accessed using standard Stokes polarimetry in photonic systems~\cite{Bleu2018}, as was demonstrated with exciton-polaritons in microcavities~\cite{Gianfrate2020, Polimeno2021, Lempicka-Mirek2022,Whittaker2021}, and with plasmonic lattices~\cite{Cuerda2024}. Beyond the case of two-band system, increasing the dimensionality of the Hilbert space leads to multiband topological phenomena, enabling the realization of a Chern insulator when polarization and sublattice degrees of freedom are combined~\cite{Karzig2015, Nalitov2015, Bleu2018, Klembt2018}, or the emergence of fragile and non-abelian topologies in systems with multiple orbitals~\cite{Po2018,Bouhon2019,Bouhon2020,Peri2020,Song2020}. So far, a robust approach for probing the sublattice pseudospin of photonic lattices, that can be extended to the multiband case, has yet to be demonstrated. Novel methods are required for accurately measuring the sublattice pseudo-spin and the eigenstates geometry in such lattices.

\begin{figure*}[t]
    \includegraphics[width=\textwidth]{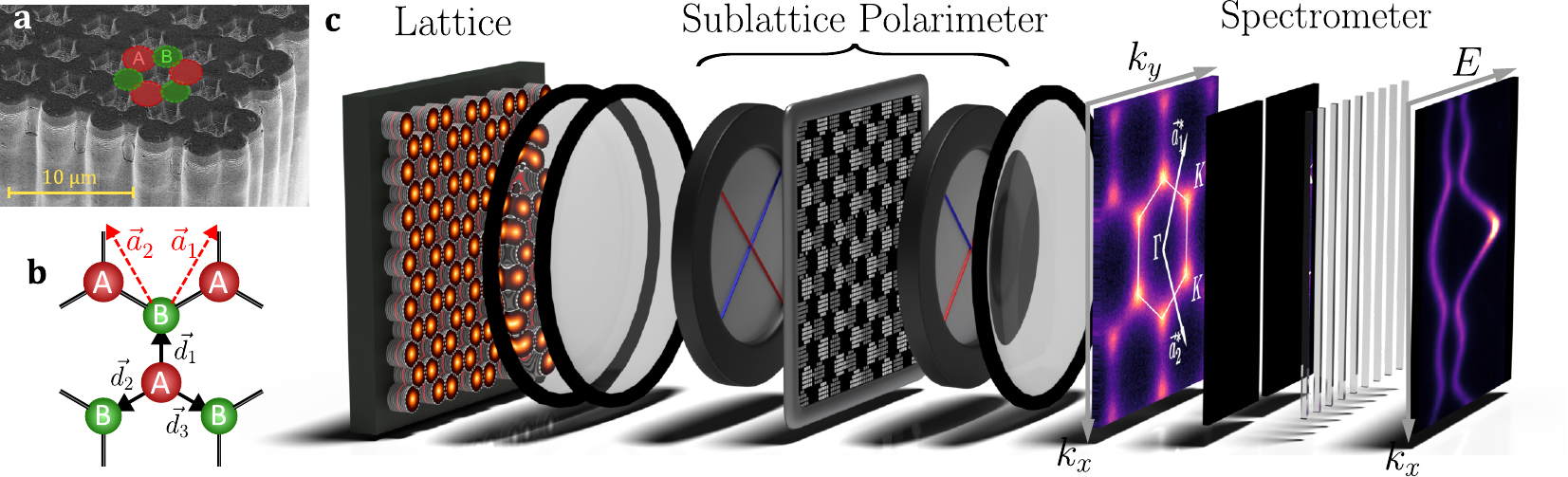}
     \caption{\textbf{a}.~Scanning electron microscope image of the staggered honeycomb lattice used in the experiment. \textbf{b}.~Schematic of the lattice showing the primitive vectors $\bm{a}_1$ and $\bm{a}_2$, as well as the $\bm{d}_1$, $\bm{d}_2$ and $\bm{d}_3$ vectors used in the definition of $\gamma(\bm{k})$ in Eq.~\ref{eq:Eq1}. \textbf{c}.~Illustration of the experimental setup. The photoluminescence emission from the lattice is directed toward the sublatice polarimeter using a pair of lenses. The polarimeter is composed of the SLM placed between two crossed polarizers. The modified emission pattern is Fourier transformed using an additional lens, and finally directed into a spectrometer for energy-momentum resolved detection on a CCD camera.}
    \label{fig:Fig1}
\end{figure*}

In this work, we realize a sublattice Stokes polarimeter for photonic lattices, which is versatile and can readily be generalized to systems with higher degrees of freedom. We demonstrate the method using photonic arrays arranged into hexagonal lattices (see Methods for details on sample structure and fabrication). The lattice features primitive vectors $\bm{a}_1$ and $\bm{a}_2$ (shown in Fig.~\ref{fig:Fig1}), and exhibits a bipartite structure composed of two sublattices, labeled $A$ and $B$ throughout this article. Within the tight-binding framework, the system implements a staggered honeycomb lattice with onsite energies $\epsilon_A$ and $\epsilon_B$ and nearest-neighbor coupling $-t$, described by the Bloch Hamiltonian:
\begin{equation}
    \hat{H}_{\bm{k}} =
    \begin{bmatrix}
        \epsilon_0 /2        & -t \gamma(\bm{k}) \\
        -t \gamma^* (\bm{k}) & -\epsilon_0 /2
    \end{bmatrix}
    = \sum_{n=0}^3 \lambda_{n}(\bm k) \hat{\sigma}_n\, ,
    \label{eq:Eq1}
\end{equation}
\noindent where $\epsilon_0=\epsilon_A-\epsilon_B$, $\gamma(\bm{k}) = e^{i\bm{k} \cdot  \bm{d}_{1}}+e^{i\bm{k} \cdot \bm{d}_{2}}+e^{i\bm{k} \cdot \bm{d}_{3}}$, and $\bm{d}_{i}$ are vectors defined in Fig.~\ref{fig:Fig1}. In Eq.~\ref{eq:Eq1}, we decompose $\hat{H}_{\bm{k}}$ onto the basis of the $\hat{\sigma}_n$ Pauli matrices, where $\lambda_0(\bm k) = 0$, $\lambda_1(\bm k) = -t\, {\rm Re} [ {\gamma (\bm k) } ] $, $\lambda_2(\bm k) = t \,{\rm Im} [ {\gamma (\bm k) } ]$ and $\lambda_3(\bm k) = \epsilon_0 / 2$.

Diagonalizing $H_{\bm{k}}$ yields two gapped energy bands $E_{n,{\bm k}}$, with $n \in \{ 1,2 \}$ the band index. The eigenstates are:
\begin{equation*}
    \ket{u_{n,\bm{k}}} =
    \begin{bmatrix}
        u_{n,{\bm{k}}}^{(A)} \\
        u_{n,{\bm{k}}}^{(B)} 
    \end{bmatrix} 
    =
    \begin{bmatrix}
        \cos(\theta_{n,\bm{k}}/2) \\
        \sin(\theta_{n,\bm{k}}/2) e^{i \phi_{n,\bm{k}}}
    \end{bmatrix} \, ,
\end{equation*}
\noindent where $\theta_{n,\bm{k}} \in [0,\pi]$ and $\phi_{n,\bm{k}} \in [0, 2\pi]$, and the first (second) vector component gives the eigenstate probability amplitude on the $A$ ($B$) sublattice. This two-level system can be mapped to a sublattice pseudo-spin on the Poincar{\'e} sphere (see Fig.~\ref{fig:Fig1}). We define the sublattice polarization Stokes vector:
\begin{equation*}
    \bm{S}_{n,{\bm{k}}} =
    \begin{bmatrix}
        S0_{n,{\bm{k}}} \\
        S1_{n,{\bm{k}}} \\
        S2_{n,{\bm{k}}} \\
        S3_{n,{\bm{k}}}
    \end{bmatrix}
    =
    \begin{bmatrix}
        1 \\
        \sin (\theta_{n,\bm{k}}) \cos (\phi_{n,\bm{k}}) \\
        \sin (\theta_{n,\bm{k}}) \sin (\phi_{n,\bm{k}}) \\
        \cos (\theta_{n,\bm{k}})
    \end{bmatrix}  \, ,
\end{equation*}
\noindent where the first component determines the Poincar{\'e} sphere radius (equal to one as the Bloch vectors are normalized), and the three other components are the cartesian coordinates of the state on the sphere. For our choice of coordinates, the north (south) pole of the Poincar{\'e} sphere corresponds to full $A$ ($B$) sublattice polarization, while states with equal amplitude on both sublattices lie on the equatorial plane.

The general idea of Stokes polarimetry for measuring the polarization state of a light beam relies on analyzing light intensity variations when passing through at least four configurations of optical elements such as polarizers and waveplates~\cite{hecht2017}. The effect of these optical elements on the polarization state of light is encoded into Mueller matrices $\mathcal{M}$, which act on the four-component Stokes vectors as ${\bm S}_{\rm out} = \mathcal{M} {\bm S}_{\rm in}$. In this work, we extend this approach to the case of a sublattice pseudo-spin. We use a reconfigurable spatial light modulator (SLM) to implement the various configurations of optical elements needed to perform the sublattice Stokes polarimetry of all eigenstates within the lattice Brillouin zone (see Fig.~\ref{fig:Fig1}).

\begin{figure*}[t]
    \includegraphics[width=\textwidth]{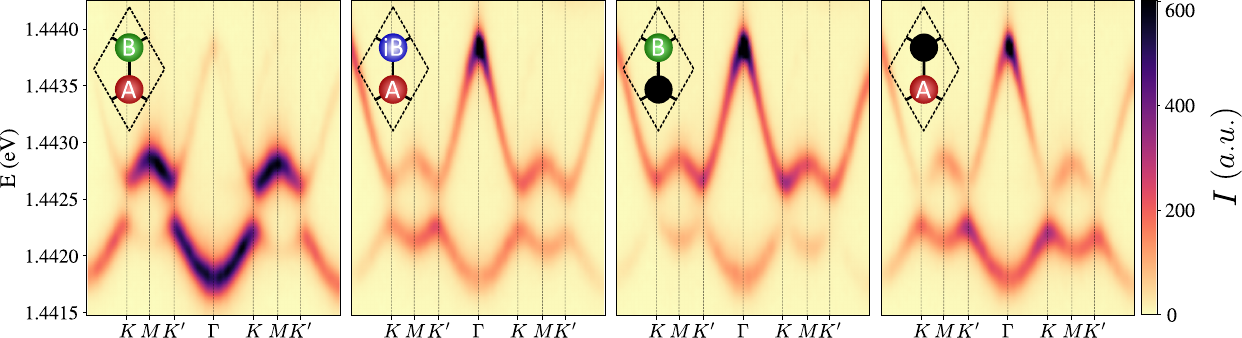}
    \caption{Band structures measured along $k_x$ ($k_y = 0$) using the four different polarimeter configurations, schematically represented in the upper left of each panel. From left to right, the panels display $I_{A+B} (k_x,E)$, $I_{A+iB} (k_x,E)$, $I_{B} (k_x,E)$ and $I_{A} (k_x,E)$}
    \label{fig:Fig2}
\end{figure*}

\begin{figure}[t]
    \includegraphics[width=0.45\textwidth]{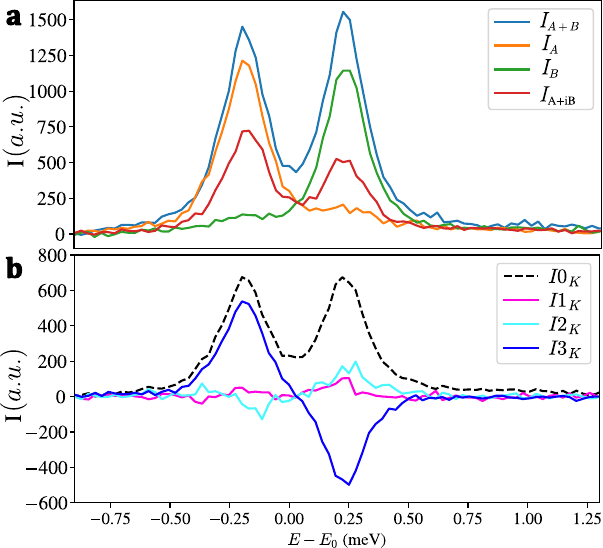}
    \caption{
    \textbf{a.}~Intensity spectra $I_{A+B}({\bm K},E)$, $I_{A+iB}({\bm K},E)$, $I_{B}({\bm K},E)$, and $I_{A}({\bm K},E)$ measured at the $K$ Dirac point. \textbf{b.}~Reconstructed intensity spectra $I_0({\bm K},E)$, $I_1({\bm K},E)$, $I_2({\bm K},E)$ and $I_3({\bm K},E)$}
    \label{fig:Fig3}
\end{figure}

Figure~\ref{fig:Fig1} provides an illustrative view of our sublattice Stokes polarimeter (see Extended Data Fig.~\ref{fig:ExtDataSetup} in Methods for details on the experimental setup). We collect the photoluminescence signal emitted through the uniformely excited lattice. We then image the lattice plane onto the SLM placed between crossed polarizer and analyzer. These components form the core of the Stokes polarimeter, as they enable (i)\quad selecting the emission originated from a small area centered on each lattice site thus enabling the mapping of the 2D continuous photonic lattice to a discrete model (see Supplementary information) and (ii)\quad imprinting a controlled phase retardation or amplitude imbalance between the fields emitted by $A$ and $B$ sites (see Methods). This enables us to implement analogues of waveplates or polarizers for the sublattice degree of freedom. Going through this setup, the sublattice Stokes vector evolves as: ${\bm S}_{n, {\bm k}}^{\rm out} = \mathcal{M} \, {\bm S}_{n, {\bm k}}$, with $\mathcal{M}$ a set of carefully designed Mueller matrices. We collect the light going through the analyzer and realize a Fourier plane image onto the entrance slit of a spectrometer coupled to a CCD camera, providing direct imaging of the band dispersion.

We first consider a configuration where the polarimeter acts as a mirror, which action on the Stokes vector is encoded into the Mueller matrix $\mathcal{M}_{A+B} = \mathds{1}$, where $\mathds{1}$ is the identity matrix. The subscript $A+B$ emphasizes that we do not alter the intrinsic phase relation between the fields emitted by both sublattices. In the tight-binding model, where each lattice site is described as a point-like emitter, each eigenmode $\ket{u_{n,\bm{k}}}$ generates the field (see Supplementary information):
\begin{equation*}
    \mathcal{\tilde{E}}_{n,\bm{k}} (E) = \mathcal{A}_{n,\bm{k}} (E) (u_{n,{\bm k}}^{(A)}+u_{n,{\bm k}}^{(B)}) e ^{-i E t / \hbar} \, ,
\end{equation*}
\noindent where $\mathcal{A}_{n,\bm{k}} (E)$ describes the photoluminescence amplitude centered around $E_{n, {\bm k}}$, with linewidth $\Gamma_{n,{\bm k}}$. The total photoluminescence intensity $I_{A+B} ({\bm k},E)= \eta \sum_{n=1}^2  | \mathcal{\tilde{E}}_{n,\bm{k}} (E) | ^2$ results from the incoherent sum of intensities in each band, and can be expressed as:
\begin{equation}
    I_{A+B} ({\bm k},E) = \sum_{n=1}^2 \eta_{n,{\bm k}} (E) ( S0_{n,{\bm k}} + p_{n,{\bm k}} S1_{n,{\bm k}}) \, ,
    \label{eq:Eq2}
\end{equation}
\noindent where $\eta_{n,{\bm k}}(E) = \eta \left| \mathcal{A}_{n,{\bm k}}(E) \right| ^2$ is the photoluminescence intensity, and $\eta$ the detection efficiency. In Eq.~\ref{eq:Eq2}, we have introduced the polarization degree $p_{n,{\bm k}}$ to account for possible experimental sources of depolarization.

Figure~\ref{fig:Fig2}(a) shows the band dispersion along $k_y = 0$  measured in this configuration of the polarimeter. The two gapped energy bands characteristic of staggered graphene are clearly visible, with the smallest energy separation occurring at the $K$ and $K'$ points. A particularly striking feature is the pronounced variation in intensity across the bands as a function of ${\bm k}$, which directly reflects the phase relationship between the fields emitted by the $A$ and $B$ sublattices. Indeed, Eq.~\ref{eq:Eq2} shows that states carrying the bonding symmetry ($S1_{n,{\bm k}} = 1$) are bright as the fields emanating from both sublattices oscillate in phase and thus interfere constructively. Conversely, modes carrying the antibonding symmetry ($S1_{n,{\bm k}} = -1$) are dark. This measurement thus provides partial information about the $S_0$ and $S_1$ components of the Stokes vector.

\begin{figure*}[t]
    \includegraphics[width=\textwidth]{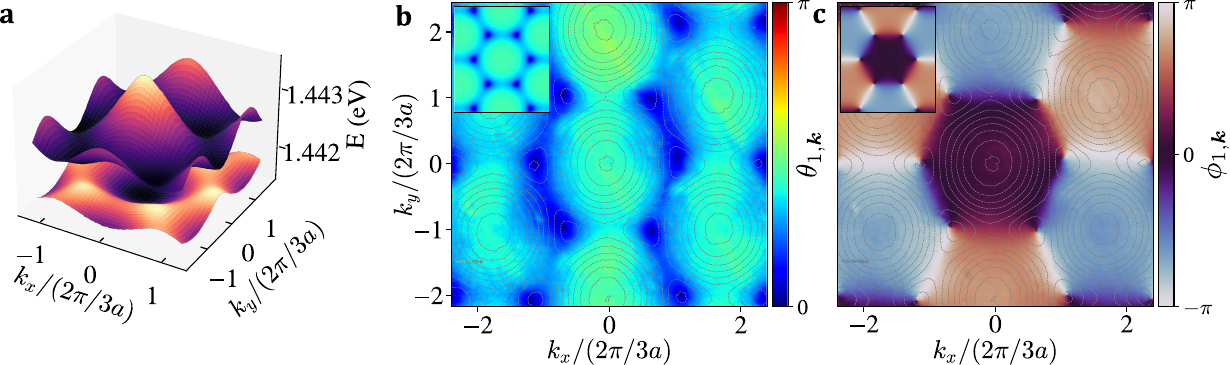}
    \caption{\textbf{a.}~Three-dimensional representation of the measured band structure, obtained by fitting for each value of $\bm{k}$ the emission lines in the $I_0({\bm k},E)$ spectrum. Recontructed \textbf{b.}~polar angle $\theta_{1,\bm{k}}$, and \textbf{c.}~azimuthal angle $\phi_{1,\bm{k}}$ for the lower band eigenvector. In \textbf{b} and \textbf{c}, faint dashed lines indicate iso-energy contours obtained from the experimentally measured dispersion. Insets display the results of tight-binding simulations, using the parameters $a = \SI{2.4}{\micro\meter}$ and $|\epsilon_0/t| = 1.26$.}
    \label{fig:Fig4}
\end{figure*}

We now use the polarimeter to modify the phase relationship and relative amplitudes between the fields in both sublattices, and perform the three complementary measurements necessary to experimentally access the full Stokes vectors of all eigenstates. Adjusting the local phases imposed by the SLM, we implement Mueller matrices acting on the Stokes vector (see Methods). Applying $\mathcal{M}_{A+iB}$ imposes a $\pi/2$ phaseshift between the two sublattices, which is equivalent to a quarter-waveplate for the sublattice pseudo-spin. $\mathcal{M}_{A}$ ($\mathcal{M}_{B}$), extinguishes the field emitted by the $B$ ($A$) sites, thus acting as a polarizer aligned along the $S_3$ axis and oriented towards the north (south) pole of the Poincar\'e sphere. The resulting photoluminescence intensities write:
\begin{equation*}
    \begin{split}
    I_{A+iB} ({\bm k},E) & = \sum_{n=1}^2 \eta_{n,{\bm k}} (E) ( S0_{n,{\bm k}}  - p_{n,{\bm k}} S2_{n,{\bm k}})/2 \, , \\
    I_{A} ({\bm k},E) & = \sum_{n=1}^2 \eta_{n,{\bm k}} (E) ( S0_{n,{\bm k}}  + p_{n,{\bm k}} S3_{n,{\bm k}})/2  \, ,  \\
    I_{B} ({\bm k},E) & = \sum_{n=1}^2 \eta_{n,{\bm k}} (E) ( S0_{n,{\bm k}}  - p_{n,{\bm k}} S3_{n,{\bm k}})/2 \, .
    \end{split}
\end{equation*}

Figures~\ref{fig:Fig2}b-d show the intensity maps measured for $k_y = 0$. Remarkably, when $\mathcal{M}_{\alpha} \neq \mathds{1}$ (Fig~\ref{fig:Fig2}b-d), the modes carrying the antibonding symmetry exhibit high intensity. This is a direct consequence of the modification of the phase relationship between the fields emanating from both sublattices. Another striking observation  in Fig~\ref{fig:Fig2} is the nearly complete extinction of the lower (upper) band intensity near both valleys in $I_{B} ({\bm k},E)$ ($I_{A} ({\bm k},E)$). This extinction is particularly visible in the linecuts shown in Fig~\ref{fig:Fig3}a, and highlights that the lower (upper) band Bloch mode is mostly localized on the $A$ ($B$) sublattice, as expected for the staggered honeycomb lattice.

Using the four measurements described above, we proceed towards the reconstruction of the Stokes parameters by calculating the following maps:
\begin{equation}
    \begin{split}
        I0 ({\bm k},E) & = I_A ({\bm k},E) + I_B ({\bm k},E) \, ,\\
        I1 ({\bm k},E) & = I_{A+B} ({\bm k},E) - I_{A} ({\bm k},E) -I_{B} ({\bm k},E) \, ,\\
        I2 ({\bm k},E) & = - 2 I_{A+iB} ({\bm k},E) + I_{A} ({\bm k},E) + I_{B} ({\bm k},E) \, ,\\
        I3 ({\bm k},E) & = I_{A}({\bm k},E) - I_{B}({\bm k},E) \, .
    \end{split}
     \label{eq:Eq3}
\end{equation}
The observable $I0 ({\bm k},E) = \sum_{n=1}^2 \eta_{n,{\bm k}} (E)  S0_{n,{\bm k}}$ is the sum of the intensities emitted through both sublattices, without interference effects. Probing this observable ensures uniform illumination across momentum space, enabling precise determination of the energy of each eigenstate. As an example, we show in Fig.~\ref{fig:Fig3}b the linecut $I0 ({\bm K},E)$. Performing double Lorentzian fits of $I0 ({{\bm k}, E})$ at each $\bm k$-point, we determine the eigenenergies $E_{1,{\bm k}}$ and $E_{2,{\bm k}}$. A three-dimensional representation of the resulting band structure is shown in Fig.~\ref{fig:Fig4}a. The three observables $I \nu ({\bm k},E)$ with $\nu \in \{ 1,2,3 \}$ isolate all three components of the Stokes vector according to $I \nu ({\bm k},E) = \sum_{n=1}^2 p_{n,{\bm k}} \eta_{n,{\bm k}} (E)  S \nu _{n,{\bm k}}$. Figure~\ref{fig:Fig3}b shows linecuts of $I \nu (K,E)$ taken at the $K$ point. We observe that $I1(K,E)$ and $I2(K,E)$ remain close to zero, while $I3(K,E)$ features two peaks with opposite signs. This measurement thus provides a quantitative estimate of the localization of $\ket{u_{1,{\bm K}}}$ ($\ket{u_{2,{\bm K}}}$) near the north (south) pole of the Poincar\'e sphere.

\begin{figure*}[t]
    \includegraphics[width=\textwidth]{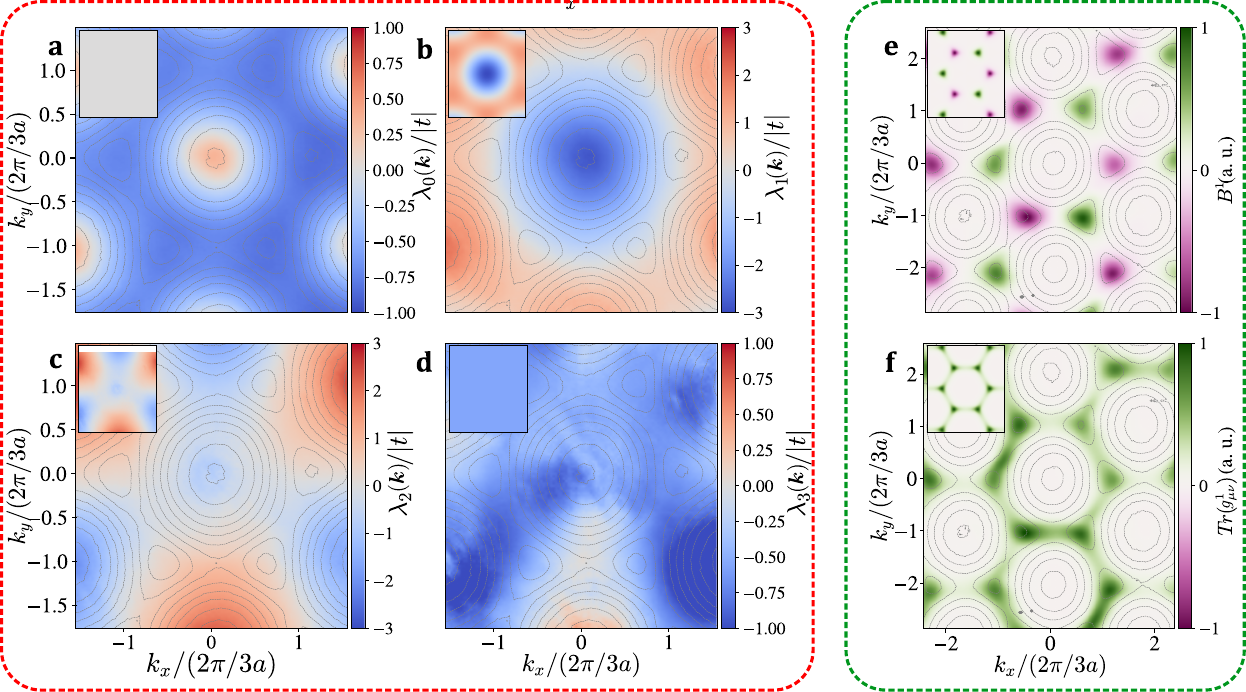}
    \caption{Experimental reconstruction of the Hamiltonian and quantum geometric tensor.
    \textbf{a-d.}~Components $\lambda_i (k_x,k_y)$ of the Hamiltonian with $i\in \{ 0,1,2,3\}$.
    \textbf{e.}~Berry curvature $B_1(k_x, k_y)$, and
    \textbf{f.}~trace of the quantum metric $\Tr [ g_{\mu \nu}^1 ({\bm k}) ]= g^1_{xx}(k_x, k_y)+g^1_{yy}(k_x, k_y)$ for the lower band. In all panels, the faint lines indicate iso-energy contours obtained from the experimentally measured dispersion. Insets display the results of tight-binding simulations, using the parameters $a = \SI{2.4}{\micro\meter}$ and $|\epsilon_0/t| = 1.26$.
    }
    \label{fig:Fig5}
\end{figure*}

Integrating $I\nu_{{\bm k},E}$ over each state, we determine the Stokes parameters $S \nu_{n,{\bm k}}$ across the Brillouin zone, as shown in Extended Data Fig.~\ref{fig:ExtDataStokes} for the lowest energy band. We then proceed to the reconstruction of the full eigenstate structure (see Methods). We show, in Fig.~\ref{fig:Fig4}b-c, the reconstructed maps of $\theta_{1, {\bm k}}$ and $\phi_{1, {\bm k}}$ for the lowest energy band, as a function of $k_x$ and $k_y$. At the center of the Brillouin zone the modes show the expected bonding symmetry ($\phi_{1,{\bm 0}} \approx 0$). At the $K$ and $K'$ points, the eigenstates are found to be fully localized at the north pole of the Poincaré sphere ($\theta_{1,{\bm K}} \approx \theta_{1,{\bm K'}} \approx 0$), corresponding to complete localization on a single sublattice. Thus, the azimuthal angle $\phi_{1,{\bm k}}$ becomes undefined, leading to singularities at $K$ and $K '$. Around these points, we clearly observe characteristic phase windings of $2 \pi$ with opposite chirality. Extended Data Fig.~\ref{fig:ExtDataEigenvectors}a-b, show $\theta_{2, {\bm k}}$ and $\phi_{2, {\bm k}}$ measured with a similar analysis for the upper band. Having determined all eigenstates, we find that the mean polarization degree of the photoluminescence signal is as high as $85 \%$, essentially limited by the polarimeter alignment. Overall, these data show very good agreement with tight-binding simulations (insets in Fig.~\ref{fig:Fig4}a-b). Slight deviations are attributed to possible experimental imperfections, which are discussed in the Supplementary Information.

Having obtained the eigenvector structure for each band, we compute their scalar product $\left| \braket{u_{1,{\bm k}}|u_{2,{\bm k}}} \right| ^2$ (see Extended Data Fig.~\ref{fig:ExtDataOrthogonality}). It stays below a few percents over the full Brillouin zone, showing that we retrieve nearly orthogonal eigenmodes. This result validates our mapping from the 2D continuous eigenmodes to Bloch eigenvectors of a discrete tight binding Hamiltonian. It also shows that possible non-hermitian effects (due for example to different losses on the $A$ and $B$ micropillars) can be safely neglected.

\begin{figure*}[t]
    \includegraphics[width=\textwidth]{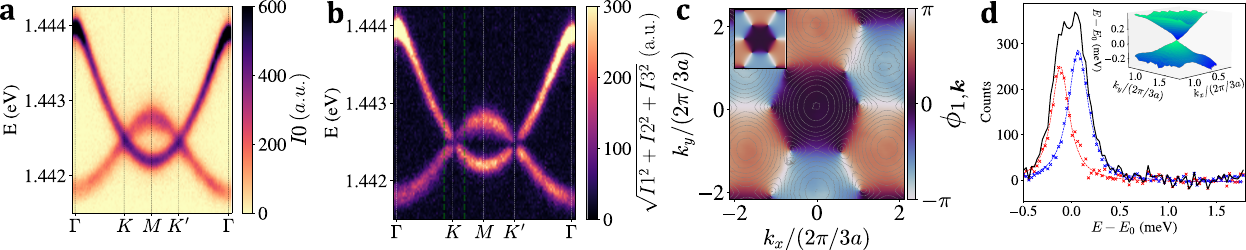}
    \caption{
    Stokes polarimetry of the gapless honeycomb lattice.
    \textbf{a}.~Measured intensity $I0(k_x,E)$ at fixed $k_y = 0$, revealing the characteristic graphene band dispersion with linear crossings at the Dirac points.
    \textbf{b}.~Measured $\sqrt{\sum_{\nu=1}^3 I\nu^2}$, indicating nearly complete depolarization of the total signal near the band touching points. 
    \textbf{c}.~Reconstructed azimuthal angle $\phi_{1,\bm{k}}$ of the lower band eigenstate.
    \textbf{d}.~Energy-resolved intensity spectra in the vicinity of $\bm{K}$. The black line shows the total intensity $I0(\bm k,E)$ for the selected value of ${\bm k}$, while the red and blue symbols indicate the two measured density matrix eigenvalues $\lambda_+$ and $\lambda_-$ as a function of energy. The red and blue solid lines are single-Lorentzian fits to the experimental data used to extract $E_{n,{\bm k}}$ with sub-linewidth resolution.}
    \label{fig:Fig6}
\end{figure*}

We now use the measured eigenvectors and eigenergies to reconstruct the system Hamiltonian. We show, in Fig.~\ref{fig:Fig5}, the measured coefficients $\lambda_i({\bm k})$ that appear in the decomposition of $\hat{H} _ {\bm k}$ into the Pauli matrices. We first focus on the nondiagonal matrix elements $\lambda_1 ({\bm k})$ and $\lambda_2 ({\bm k})$ in Fig.~\ref{fig:Fig5}b-c. Near the center of the first Brillouin zone, we see that the nondiagonal coupling coefficient is purely real ($\lambda_2 (\Gamma) \simeq 0$), which leads to eigenstates carrying the bonding or anti-bonding symmetry. Conversely, near the $K$ and $K'$ points, the nondiagonal coupling coefficients cancel ($\lambda_{1,2} (K) \simeq \lambda_{1,2} (K') \simeq 0$), reflecting sublattice localization of the eigenstates. Focusing on the diagonal matrix elements, we notice in Fig.~\ref{fig:Fig5}d that $\lambda_3 ({\bm k}) $ is overall positive over the full Brillouin zone. This term corresponds to the onsite energy we have engineered to implement gapped graphene. The results of tight-binding calculations are displayed in the insets, and show good agreement for $\lambda_1({\bm k})$, $\lambda_2({\bm k})$ and $\lambda_3({\bm k})$. Small, unexpected variations in $\lambda_3({\bm k})$ are attributed to experimental imperfections.

Interestingly, Fig~\ref{fig:Fig5}a reveals significant deviations of the $\lambda_0 ({\bm k}) $ coefficient, which takes non-zero values near the center of the Brillouin zone, contrarily to what is expected according to Eq.~\ref{eq:Eq1}. The observed modulation of $\lambda_0 ({\bm k})$ is the signature of a non-zero next-nearest neighbor coupling $t'$ in the Hamiltonian, which we precisely estimate from the measured modulation amplitude of $\lambda_0 ({\bm k})$. Including this additional coupling in numerical tight-binding simulations, we obtain good quantitative agreement with the experimental data (see Extended Data Fig.~\ref{fig:ExtDataHamiltonian}). We note that the observed asymmetry of two lowest energy bands with respect to the center of the gap is also a direct consequence of this non zero value of $t'$~\cite{CastroNeto2009,Jacqmin2014}.
In Ref.~\cite{Mangussi2020}, the physical origin of this distortion of the two bands formed by the hybridization of the $S$-orbitals, is shown to be related to their coupling to higher energy states ($P$-orbitals), an effect that is effectively captured by introducing a second-neighbor coupling in the two-band model.

Another direct application of the measurement of the lattice eigenstate structure is the reconstruction of the quantum geometric tensor $Q_{\mu \nu}^n$:
\begin{equation*}
    Q_{\mu \nu}^n ({\bm k}) = \bra{\partial_{k_\mu} u_{n,{\bm k}}} \left(\mathds{1} - \dyad{u_{n,{\bm k}}}{u_{n,{\bm k}}} \right) \ket{\partial_{k_\nu} u_{n,{\bm k}}} \, ,
\end{equation*}
\noindent which decomposes into real and imaginary parts as follows:
\begin{equation*}
    Q_{\mu\nu}^n({\bm k}) =
    g_{\mu \nu}^n ({\bm k}) - \frac{i}{2} \Omega_{\mu \nu}^n ({\bm k}) \, ,
\end{equation*}
\noindent where $g_{\mu \nu}^n ({\bm k})$ is the quantum metric and $B_n ({\bm k}) = \Omega_{x y}^n ({\bm k})$ is the Berry curvature for band $n$. Figure~\ref{fig:Fig5}e, displays the reconstructed $B_1 ({\bm k})$ versus $k_x$ and $k_y$. We observe that $B_1 ({\bm k})$ is zero in most parts of the Brillouin zone, except in the vicinity of the $K$ and $K'$ points, where it peaks due to the winding of $\phi_{1,{\bm k}}$. We compute the valley Chern numbers $\mathcal{C}_n(K)$ and $\mathcal{C}_n(K')$ defined as the integrals of $B_n ({\bm k})$ around surfaces $\Sigma_K$ and $\Sigma_{K'}$ centered around $K$ or $K'$, and each covering one half of the Brillouin zone :
\begin{equation*}
    \mathcal{C}_n(\nu) = \frac{1}{2 \pi} \iint _ {\Sigma_\nu} B_n ({\bm k}) d^2 {\bm k} \, ,
\end{equation*}
\noindent where $\nu \in \{ K, K'\}$. We obtain $\mathcal{C}_1(K) = -0.25 \pm 0.01$, and $\mathcal{C}_1(K') = 0.25 \pm 0.01$. For comparison, tight-binding simulations using the Hamiltonian and lattice parameters specified in Eq.~\ref{eq:Eq1} give the theoretical value $\mathcal{C}^{\rm th}_1(K) = -0.28$, thus highlighting the high accuracy of our measurement protocol. As the contributions from both valleys have equal magnitude and opposite sign, the total Chern number vanishes $C_1 = 0.00 \pm 0.01$, as expected for a valley-Chern insulator with conserved time-reversal symmetry.

We show, in Extended Data Fig.~\ref{fig:ExtDataQGT}, the other components of the quantum geometric tensor, which all show very good agreement with the simulations. One physical quantity of interest to quantify the distance between eigenstates within a single band and how fast they evolve in momentum space is the trace of the quantum metric $\Tr [ g_{\mu \nu}^1 ({\bm k}) ]$. We display, in Fig.~\ref{fig:Fig5}f the experimentally measured $\Tr [ g_{\mu \nu}^1 ({\bm k}) ]$, and observe that it is peaked around the $K$ and $K'$ valleys, in agreement with the tight-binding simulations shown in the inset. Overall our method enables us to fully characterize the geometry of the space spanned by the eigenstates and its topology.

So far, we have implemented our method to determine the Bloch eigenmodes of gapped bipartite lattices. However, many fascinating physical phenomena connected to Dirac physics, topological insulators, and Euler topology occur when bands closely overlap or even touch. Experimentally accessing this physics requires sub-linewidth spectroscopic tools. In the final part of this article, we show that our sublattice Stokes polarimeter can be extended beyond the case of gapped systems, provided the orthogonality of the Bloch Hamiltonian eigenmodes is assumed.

We apply our four measurement protocol to a gapless honeycomb lattice composed of identical micropillar cavities (see Methods), and reconstruct the intensity maps $I \nu ({\bm k}, E)$. Fig.~\ref{fig:Fig6}a shows $I 0 ({\bm k}, E)$, where we observe the characteristic graphene dispersion featuring two bands that touch at the $K$ and $K'$ points. When the emissions from both eigenmodes spectrally overlap, the Stokes vector of the detected signal corresponds to the sum of the individual Stokes vectors. For orthogonal eigenmodes, the individual Stokes vectors are diametrically opposed on the Poincaré sphere, so that all three polarization components of the total Stokes vector simultaneously vanish. We note that non-Hermitian effects in the system could cause the eigenmodes to become non-orthogonal near the band touching points~\cite{Bergholtz2021,Su2021}. Nevertheless, we anticipate these effects to be negligible as all lattice sites are identical so that the imaginary part of the Hamiltonian is proportional to the identity matrix and the eigenvectors coincide with those of the Hermitian part of the Hamiltonian. We experimentally test the orthogonality of the Bloch vectors by plotting, in Fig.~\ref{fig:Fig6}b, the measured $\sqrt{ \sum_{\nu=1}^3 I \nu ({\bm k},E) ^2 }$. We observe that this quantity approaches zero close to the $K$ and $K'$ points, where both eigenmodes overlap. This observation validates experimentally the orthogonality of the Bloch vectors in this lattice.

We extend our data analysis to the regime where bands overlap, and take into account the mixed nature of the detected signal. For each ${\bm k}$ and $E$, we construct a density matrix representing the mixed state, as described in the Methods. Diagonalizing this matrix yields eigenvectors corresponding to the measured Bloch eigenstates, and eigenvalues that provide the spectral weights of these states (see Methods). The resulting $\phi_{1,{\bm k}}$, shown in Fig.~\ref{fig:Fig6}c, clearly reveals, with high resolution, the $2 \pi$ phase windings expected around the Dirac cones. Moreover, by plotting the density matrix eigenvalues versus energy, we reconstruct the individual spectral lines. We show, in Fig.~\ref{fig:Fig6}d, the spectra obtained near a Dirac point, where the initially broad line is clearly separated into two distinct contributions. Single-lorentzian fits of these two lines enable accessing the values of the individual $E_{n,\bm{k}}$ with sublinewidth precision. The three-dimensional representation of the band dispersion, shown in Extended Data Fig.~\ref{fig:ExtDataGraphene}, faithfully captures the entire band dispersion and accurately resolves the Dirac cones.

To conclude, we report on the design and implementation of a sublattice Stokes polarimeter that enables momentum-resolved reconstruction of Bloch eigenstates in photonic lattices, with sub-linewidth precision. This approach allows to directly accessing the full system Hamiltonian and extract parameters that are inaccessible through band structure measurements alone. This is especially relevant near band degeneracies, where rich topological eigenstate structures are expected to emerge~\cite{Karzig2015, Nalitov2015}. We anticipate that extending our work to engineered non-Hermitian photonic systems is an exciting avenue, since our Stokes polarimeter could enable  high precision investigation of non orthogonal eigenstates together with the emergence of spectral features such as exceptional points and Fermi arcs~\cite{Gao2015,Ghatak2019,Bergholtz2021,Su2021,Liao2021,Hu2025}. Going beyond the current two-band implementation, this technique can straightforwardly be extended to multi-band systems involving higher orbital or other spinor degrees of freedom such as the polarization of light, unlocking experimental studies of multi-band topology, Euler curvature and non-Abelian topology~\cite{Karzig2015, Nalitov2015, Bleu2018, Klembt2018, Po2018, Bouhon2019, Bouhon2020, Peri2020, Song2020}.


%


\section{Methods}

\subsection{Sample fabrication}

The samples are fabricated from an epitaxially grown AlGaAs semiconductor microcavity consisting of two distributed Bragg reflectors (DBRs) made of $\mathrm{Al}_{0.95}\mathrm{Ga}_{0.05}\mathrm{As}$ and $\mathrm{Al}_{0.1}\mathrm{Ga}_{0.9}\mathrm{As}$, with 26 (top) and 30 (bottom) mirror pairs, and a central GaAs spacer of optical thickness $\lambda_0 = \SI{864}{\nano\meter}$. A single $\SI{17}{\nano\meter}$-thick $\mathrm{In}_{0.05}\mathrm{Ga}_{0.95}\mathrm{As}$ quantum well is embedded at the antinode of the cavity field. The exciton resonance, centered at \SI{1.450}{\electronvolt} at \SI{4}{\kelvin}, is strongly coupled to the cavity mode with a Rabi coupling $\Omega_R = \SI{3.6}{\milli\electronvolt}$, resulting in exciton-polaritons as the elementary optical excitations. The exciton is $\SI{9.7}{\milli\electronvolt}$ detuned above the $\bm{k}=0$ cavity mode, yielding predominantly photonic bands. The exciton primarily enables relaxation into the polariton bands under non-resonant excitation.

Micropillar lattices are defined using electron-beam lithography followed by dry etching. The onsite energies $\epsilon_A$ and $\epsilon_B$ are tuned via the pillar diameters, while the nearest-neighbor coupling $-t$ is set by the interpillar spacing. For the staggered honeycomb lattice, we use radii $R_A = \SI{1.57}{\micro\meter}$ and $R_B = \SI{1.27}{\micro\meter}$ with a lattice constant $a = \SI{2.4}{\micro\meter}$, yielding an onsite energy contrast $\epsilon_0 = \epsilon_A - \epsilon_B \simeq \SI{-0.426}{\milli\electronvolt}$ and $t = \SI{0.337}{\milli\electronvolt}$. For the gapless honeycomb lattice, identical pillars ($R_A = R_B = \SI{1.425}{\micro\meter}$) are used, with the same lattice constant.

\subsection{Experimental setup}

A detailed schematic of the experimental setup is shown in Extanded Data Fig.~\ref{fig:ExtDataSetup}. Optical excitation is provided by a continuous-wave laser tuned to $\SI{1.589}{\electronvolt}$ ($\sim \SI{780}{\nano\meter}$), far blue-detuned from the polariton modes probed in the experiment. The samples are maintained at low temperature ($\SI{4}{\kelvin}$) using a closed-cycle cryostation. The photoluminescence signal is collected in transmission using an aspheric lens of numerical aperture $\rm{N.A. = 0.55}$ and focal length $f_c = \SI{4.56}{\milli\meter}$ placed inside the cryostat.

The sample plane is then imaged onto a Hamamatsu Spatial Light Modulator (SLM) using three convergent lenses of focal lengths $f_1= \SI{200}{\milli\meter}$, $f_2= \SI{200}{\milli\meter}$, and $f_3= \SI{400}{\milli\meter}$. The magnification is adjusted so that each micropillar image covers approximately $10 \times 10$~pixels on the SLM. This imaging condition is essential as it enables us to select spatially the central region of each micropillar with sub-pillar size resolution. This is a crucial requirement for valid mapping of the 2D continuous photonic lattice to a discrete tight-binding model (see Supplementary information).

To perform the Stokes polarimetry of the Bloch eigenvectors, both amplitude and phase control of the wavefront are required in the image plane of the lattice. To this end, a polarizer oriented at $45^\circ$ relative to the SLM optical axis defines the input polarization. The light reflected from the SLM is deflected by a small angle ($\theta \lesssim 5^\circ$) and passes through an analyzer oriented orthogonally to the input polarizer, enabling the SLM to operate in a combined phase-and-amplitude modulation mode.

Finally, a system of three additional lenses ($f_4= \SI{500}{\milli\meter}$, $f_5= \SI{100}{\milli\meter}$, and $f_6= \SI{300}{\milli\meter}$), enables imaging the Fourier plane of the first lens (corresponding to momentum-resolved emission) onto on the entrance slit of a spectrometer coupled to an Andor Solis $1024 \times 1024$~CCD with $\SI{13.3}{\micro\meter} \times \SI{13.3}{\micro\meter}$ pixel size. The last lens is mounted on a translation stage, allowing measuring the band dispersion versus $k_x$ at any $k_y$ value. The overall spectral resolution is $\delta E = \SI{28}{\micro\electronvolt\per\pixel}$.

We emphasize that a precise alignment between the phase pattern displayed on the SLM and the real-space emission of the lattice is critical for accurate reconstruction of the eigenstates. To achieve this alignment, we illuminate the sample with a non-resonant pump beam, and increase the power to induce lasing in a well-defined mode. We image, with high resolution, the spatial profile of the lasing mode after passing through the Stokes polarimeter and accurately superimpose the mask pattern displayed on the SLM with the emission pattern of the bright lasing mode. This alignment procedure is essential to ensure the fidelity of the eigenstate reconstruction and to minimize artifacts arising from mask misalignment (see Supplementary Information).

\subsection{Polarimeter operation and implementation of Mueller matrices for the sublattice pseudo-spin}

The polarizer placed in front of the SLM projects the photoluminescence signal emitted by both sublattices onto the diagonal polarization state $\ket{D}$, oriented at $45^{\circ}$ relative to the SLM optical axis (aligned along the horizontal direction). Each pixel of the SLM introduces a voltage-controlled, spatially varying phase shift $\varphi (x_{\rm SLM}, y_{\rm SLM} )$ to the horizontal component of the field $\ket{H}$, while leaving the vertical component $\ket{V}$ unchanged, where $x_{\rm SLM}$ and $y_{\rm SLM}$ are the wavefront coordinates in the SLM plane. Finally, the analyzer projects the reflected light going through the polarimeter onto the antidiagonal state $\ket{AD}$, so that the final polarization state of the light that has reflected off pixel $(x,y)$ is
\begin{equation*}
    \ket{f_{x, y}} = i \sin(\varphi_{x, y}/2) e ^ {  i \varphi_{x, y} /2 } \ket{AD} \, .
\end{equation*}
This setup thus implements a complex transfer function $F(\varphi_{x,y})$ acting locally on the wavefront according to:
\begin{equation*}
    F(\varphi_{x,y}) = i \,\sin(\varphi_{x,y}/2)\,e^{i\varphi_{x,y}/2} \, ,
\end{equation*}
highlighting the local control achieved simultaneously over amplitude and the phase, as required for performing our four measurements. The transfer function is shown in Extanded Data Fig.~\ref{fig:ExtDataSLMcalib}a. We notice in particular the light reflecting off pixels such that $\varphi_{x,y} = 0$, is fully extinguished by the analyzer.

The SLM is controlled by uploading grayscale images, where each gray level (ranging from 0 to 255) implements a specific value of $\varphi_{x,y}$. We calibrate the response of the SLM by measuring the transmitted intensity as a function of gray level for a uniform input mask (see Extended Data Fig.~\ref{fig:ExtDataSLMcalib}b), which allows us to determine the correspondence between gray levels and $\varphi_{x,y}$ (see Extended Data Fig.~\ref{fig:ExtDataSLMcalib}c).The phase masks used in the experiment are shown in Extended Data Fig.~\ref{fig:ExtDataMasks}, where the black pixels ($\varphi_{x,y} = 0$), trace regions that are filtered out by the polarimeter. The remaining pixels (gray levels) correspond to small circular regions centered on each pillar, which are assigned nonzero values of $\varphi_{x,y}$. This allows us to selectively collect and shape the emission from the center of each lattice site.

We derive, in the Supplementary Information, the mapping between the 2D photonic lattice and the tight-binding model, where we take into account the continuous nature of the photonic wavefunctions extending over the micropillar array. A key requirement for the mapping to be valid is that the phase of the wavefunction remains approximately constant within the circular regions selected using the SLM. This imposes a constraint on the radius $R_m$ of the circular regions, which must be sufficiently small to suppress phase variations arising from the underlying continuous eigenmodes, while remaining large enough to keep a sufficient level of signal (see Supplementary Information).
In the experiments presented here, we chose $R_m = \SI{0.71}{\micro \meter}$, a value that ensures accurate reconstruction of the Bloch eigenstates across the first Brillouin zone. Deviations from this condition may lead to experimental imperfections, which are discussed in the Supplementary Information.

The SLM mask used for measuring $I_{A+B} ({\bm k,E})$ is shown in Extended Data Fig.~\ref{fig:ExtDataMasks}, the circular regions centered on each pillar feature a grey level corresponding to $\varphi = \pi$ both for the $A$ and $B$ sublattices. This configuration preserves the phase relationship between both sublattices, without any signal attenuation. Therefore, the Mueller matrix corresponding to this operation is the identity matrix:
\begin{equation*}
    \mathcal{M}_{A+B} =
    \begin{pmatrix}
        1 & 0 & 0 & 0  \\
         0 & 1 & 0 & 0 \\
         0 & 0 & 1 & 0 \\
         0 & 0 & 0 & 1 
    \end{pmatrix} \, .
\end{equation*}

For measuring $I_{A} ({\bm k,E})$, we select $\varphi = \pi$ for the regions centered on the $A$ pillars while $\varphi = 0$ for the $B$ pillars (see Extended Data Fig.~\ref{fig:ExtDataMasks}). Therefore all the light coming from the $A$ sublattice is transmitted, while the rest is completely extinguished. For measuring $I_{B} ({\bm k,E})$, the situation is reversed, where only the light coming from the $B$ sublattice is fully transmitted, while the rest is extinguished. In this configuration, the SLM acts as a polarizer for the sublattice degree of freedom, aligned along the $S_3$ axis and oriented towards the north or south pole of the Poincaré sphere. The corresponding Mueller matrices write:
\begin{align*}
    \mathcal{M}_{A} = \frac{1}{2}
    \begin{pmatrix}
         1 & 0 & 0 & 1  \\
         0 & 0 & 0 & 0 \\
         0 & 0 & 0 & 0 \\
         1 & 0 & 0 & 1 
    \end{pmatrix} && 
    \mathcal{M}_{B} = \frac{1}{2}
    \begin{pmatrix}
         1 & 0 & 0 & -1  \\
         0 & 0 & 0 & 0 \\
         0 & 0 & 0 & 0 \\
         -1 & 0 & 0 & 1 
    \end{pmatrix}
\end{align*}

Finally, for measuring $I_{A+iB} ({\bm k,E})$, we chose $\varphi = \pi/2$ for all regions centered on the $A$ sites and $\varphi = -\pi/2$ for all regions centered on the $B$ sites. Overall, this attenuates the field amplitudes on both sublattices by $1/\sqrt{2}$, and introduces a $\pi / 2$ phaseshift between the field emitted by both sublattices. This is equivalent to the combined action of a $50 \% $ attenuation filter and a quarter waveplate for the sublattice degree of freedom leading to the Mueller matrix:
\begin{equation}
    \mathcal{M}_{A+iB} = \frac{1}{2}
    \begin{pmatrix}
        1 & 0 & 0 & 0  \\
         0 & 0 & - 1 & 0 \\
         0 &  1 & 0 & 0 \\
         0 & 0 & 0 & 1 
    \end{pmatrix} \, .
\end{equation}

Considering the input Stokes vector (before the polarizer):
\begin{equation*}
    \bm{S}^{\rm in}_{n,{\bm{k}}} =
    \begin{bmatrix}
        S0_{n,{\bm{k}}} \\
        p_{n,{\bm k}} S1_{n,{\bm{k}}} \\
        p_{n,{\bm k}} S2_{n,{\bm{k}}} \\
        p_{n,{\bm k}} S3_{n,{\bm{k}}}
    \end{bmatrix}
    \, ,
\end{equation*}
the output Stokes vector (after the analyzer) is: $\bm{S}^{\rm out}_{n,{\bm k}} = \mathcal{M} \bm{S}^{\rm in}_{n,{\bm k}}$. Finally, the total intensity in band $n$ can be directly expressed as a function of the Stokes components of $\bm{S}^{\rm out}_{n,{\bm k}}$ (see Eq.~\ref{eq:Eq2}):
\begin{equation*}
    I_{n, \rm out} (\bm{k}, E) = \eta_{n,{\bm k}}(E) (S0_{n,{\bm k}}^{\rm out} + S1_{n,{\bm k}}^{\rm out}) \, .
\end{equation*}
As a consequence, we find:
\begin{equation*}
    \begin{split}
    I_{n, A+B} ({\bm k},E) & = \eta_{n,{\bm k}} (E) ( S0_{n,{\bm k}}  + p_{n,{\bm k}} S1_{n,{\bm k}}) \, , \\
    I_{n, A} ({\bm k},E) & = \eta_{n,{\bm k}} (E) ( S0_{n,{\bm k}}  + p_{n,{\bm k}} S3_{n,{\bm k}})/2  \, ,  \\
    I_{n, B} ({\bm k},E) & = \eta_{n,{\bm k}} (E) ( S0_{n,{\bm k}}  - p_{n,{\bm k}} S3_{n,{\bm k}})/2 \, , \\
    I_{n, A+iB} ({\bm k},E) & = \eta_{n,{\bm k}} (E) ( S0_{n,{\bm k}}  - p_{n,{\bm k}} S2_{n,{\bm k}})/2 \, .
    \end{split}
\end{equation*}

\subsection{Reconstruction of the Bloch eigenstates from the measured Stokes parameters}

\textbf{\textit{Staggered honeycomb lattice.}} In the case of the staggered honeycomb lattice, the bands are well separated in energy, and we are able to address each state individually. By integrating $I \nu ({\bm k},E)$ over an energy window $\Delta E_{n,{\bm k}}$ spanning the emission line of a given mode, we determine $S \nu_{n,{\bm k}}$ for each band individually across the entire Brillouin zone. We then reconstruct the full eigenstate structure by computing the following quantities:
\begin{equation*}
    \begin{aligned}
    S \nu_{n,{\bm k}} &= \dfrac{ \int_{\Delta E _{n,{\bm k}}} I \nu ({\bm k},E)\, dE }{ \int_{\Delta E _{n,{\bm k}}} \sqrt{ \sum_{\mu = 1}^3 I \nu ({\bm k},E)^2} \, dE } \, , \quad \nu \in \{ 1,2,3 \} \, , \\
    \theta_{n, {\bm k}} &= \arccos (S3_{n, {\bm k}}) \, , \\
    \phi_{n, {\bm k}} &= \atan2 (S2_{n, {\bm k}},S1_{n, {\bm k}}) \, , \\
     p_{n, {\bm k}} &= \dfrac{ \int_{\Delta E _{n,{\bm k}}} \sqrt{\sum_{\mu = 1}^3 I \nu ({\bm k},E)^2}\, dE }{  \int_{\Delta E _{n,{\bm k}}} I0 ({\bm k},E)\, dE  } \, ,
    \end{aligned}
\end{equation*}
\noindent where $\atan2$ is the 2-argument arctangent.

\bigskip

\textbf{\textit{Gapless honeycomb lattice.}} In the case of the gapless honeycomb lattice, we generalize the data analysis to ${\bm k}$ values where the bands overlap. To take into account the mixed nature of the detected signal, we construct, for each ${\bm k}$ and $E$, the following density matrix:
\begin{align*}
    \hat{\rho}(\bm{k},E) & = \frac{1}{2} \sum_{\nu =0}^{3} I\nu(\bm{k},E) \hat{\sigma}_\nu \\
    & = \frac{1}{2} \sum_{n=1}^2 \eta_{n,\bm{k}} (E) \left( \hat{\sigma}_0 + p_{n,\bm{k}} \sum_{\nu=1}^3 S \nu_{n, {\bm k}} \hat{\sigma}_\nu \right) \, .
\end{align*}
By construction of the density matrix, we find regimes where one or both eigenvectors of $\hat{\rho}(\bm{k},E)$ coincide with $\ket{u_n({\bm k})}$. For instance, when the bands are well separated in energy, the density matrices associated with band $n$ have eigenstates $\ket{u_n({\bm k})}$ and $\ket{u_n({\bm k}) ^ \perp}$, with eigenvalues $\eta_{n,{\bm k}}(E) (1 \pm p_{n,{\bm k}})/2$. When the two bands spectrally overlap, the situation becomes more intricate. Nevertheless, having verified that the two Stokes vectors ${\bm S}_{1,{\bm k}}$ and ${\bm S}_{2,{\bm k}}$ point in diametrically opposite directions ($S\nu_{2,{\bm k}} = -S\nu_{1,{\bm k}}$ for $\nu \in \{ 1, 2,3 \}$), $\hat{\rho}(\bm{k},E)$ simplifies to:
\begin{align*}
    \hat{\rho}(\bm{k},E) = \overline{\eta}_{\bm k} (E) \hat{\sigma}_0 + \delta {\eta}_{\bm k} (E) \sum_{\nu=1}^3 S \nu _{1, {\bm k}} \hat{\sigma}_\nu  \, ,
\end{align*}
where  $\overline{\eta}_{\bm k} (E) = \left[ \eta_{1, {\bm k}}(E) + \eta_{2, {\bm k}}(E) \right] / 2$ is the average intensity, $\delta {\eta}_{\bm k} (E) = \left[ \eta_{1, {\bm k}}(E) p_{1, {\bm k}} - \eta_{2, {\bm k}}(E) p_{2, {\bm k}} \right] /2$. In this case the two orthogonal eigenstates of  $\hat{\rho}(\bm{k},E)$ coincide with the lattice Bloch eigenstates:
\begin{equation*}
    \hat{\rho}(\bm{k},E) = \lambda_+ (\bm{k},E) \ketbra{u_{1, {\bm k}}}{u_{1, {\bm k}}} + \lambda_- (\bm{k},E) \ketbra{u_{2, {\bm k}}}{u_{2, {\bm k}}} \, ,
\end{equation*}
with $\lambda_{\pm} (\bm{k},E) = \overline{\eta}_{\bm k} (E) \pm \delta {\eta}_{\bm k} (E)$. To reconstruct the eigenvectors $\ket{u_n(\bm{k})}$ from the measurements, we jointly diagonalize \cite{Shi2011} the set of density matrices associated with a given $\bm{k}$, restricting to energy values where the total intensity $I_0(\bm{k},E)$ exceeds a given threshold, and determine the measured Bloch eigenvectors. Interestingly, owing to the high polarization degree observed in our lattices, the density matrix eigenvalues yield $\lambda_+ (\bm{k},E) \simeq \eta_{1, {\bm k}}(E)$, $\lambda_- (\bm{k},E) \simeq \eta_{2, {\bm k}}(E)$. This enables us to reconstruct the intensity profile of each mode with sublinewidth resolution, by evaluating $\braket{u_{n, \bm{k}} | \hat{\rho}(\bm{k},E) | u_{n, \bm{k}}}$.

\subsection{Computation of the quantum geometric tensor}

An efficient method to compute the Berry curvature $\mathcal{B}_n({\bm k})$ at a given momentum ${\bm k}$, introduced in Ref.~\cite{Fukui2005}, relies on computing the product of wavefunction overlaps around elementary plaquettes in the Brillouin zone. Specifically, we define the four states at the corners of an elementary plaquette as follows: $\ket{u1} = \ket{u_{n,\bm{k}}}$, $\ket{u2} = \ket{u_{n, \bm{k} + dk_x \mathbf{e_x}} }$, $\ket{u3} = \ket{u _ {n, \bm{k}+dk_x \mathbf{e_x}+dk_y \mathbf{e_y} }}$, $\ket{u4} = \ket{u_{n,\bm{k}+dk_y \mathbf{e_y}}}$. Expanding the exact formula of the Berry curvature to first order, one obtains:
\begin{equation*}
    B_n(\bm{k}) dk_x dk_y = \arg \left( \braket{u1|u2} \braket{u2|u3} \braket{u3|u4} \braket{u4|u1} \right) \, .
\end{equation*}
\noindent We generalize this approach to all components of the quantum geometric tensor, and in particular to the quantum metric:
\begin{align*}
    g^n_{xx}(\bm{k}) & = \left( 1-\braket{u1|u2} \right) / (dk_x dk_x) \, , \\
    g^n_{yy}(\bm{k}) & = \left( 1-\braket{u1|u4} \right) / (dk_y dk_y) \, , \\
    g^n_{xy}(\bm{k}) & = \left( \braket{u1|u2} + \braket{u1|u4} - \braket{u1|u3} - 1 \right) / (2 dk_x dk_y) \, .
\end{align*}
\noindent We note that these expressions rely on scalar products between Bloch eigenvectors corresponding to neighboring $\bm k$-points, separated by a step $dk$ determined by the pixel resolution of the CCD camera in momentum-space imaging. To reduce noise in the measurements, we apply a Gaussian smoothing filter of width $\sigma = 2$ pixels to the maps of Stokes parameters used to reconstruct the eigenvectors.The smoothing introduces a source of uncertainty in the measurement of quantum geometric tensor, of the order of a few percents, which represents the dominating error contribution in the determination of the valley-Chern numbers.

\section*{Data availability}

All datasets generated and analysed during this study are available upon request from the corresponding author.

\section*{Code availability}

All codes generated during this study are available upon request from the corresponding author.


%


\clearpage

\section*{Acknowledgments}

We thank Alberto Amo and Jean-No\"el Fuchs for insightful discussions, and Tom Guillemot for his contributions to the SLM setup during the early stages of the project.

This work was primarily supported by the European Research Council (ERC) under the European Union's Horizon 2020 research and innovation programme through the Starting Grant ARQADIA (grant agreement no. 949730). Additional support was provided by the Paris Ile-de-France R\'egion via DIM SIRTEQ and DIM QUANTIP, the RENATECH network and the General Council of Essonne, as well as by the ERC under the Horizon Europe programme (project ANAPOLIS, grant agreement no. 101054448).


\section*{Author contributions}

M.G. and N.P. proposed the original idea of modifying interference conditions by acting on the sublattices. M.G. built the experimental setup and conducted the experiments. Both M.G. and C.B. performed the data analysis, and developed the codes used for data processing and numerical simulations. S.R. designed the sample structure and coordinated the sample fabrication together with M.M. and A.L. for the molecular beam epitaxy growth, M.G. for the lithography mask design, L.L.G. for the electron-beam lithography, and A.H. and I.S. for etching the structures. M.G. prepared the figures for the manuscript. S.R., M.G., J.B., and C.B. wrote the manuscript. S.R. conceived and supervised the project. S.R., J.B., M.G. and C.B., took part in scientific discussions throughout the project.


\section*{Competing interests}

The authors declare no competing interests.


\section*{Additional information}

\textbf{Correspondence and requests for materials} should be addressed to Sylvain Ravets.


\onecolumngrid

\setcounter{figure}{0}
\makeatletter
\renewcommand{\fnum@figure}{Extended Data Fig.~\thefigure}
\makeatother

\begin{figure*}[p]
    \centering
    \includegraphics[width=\textwidth]{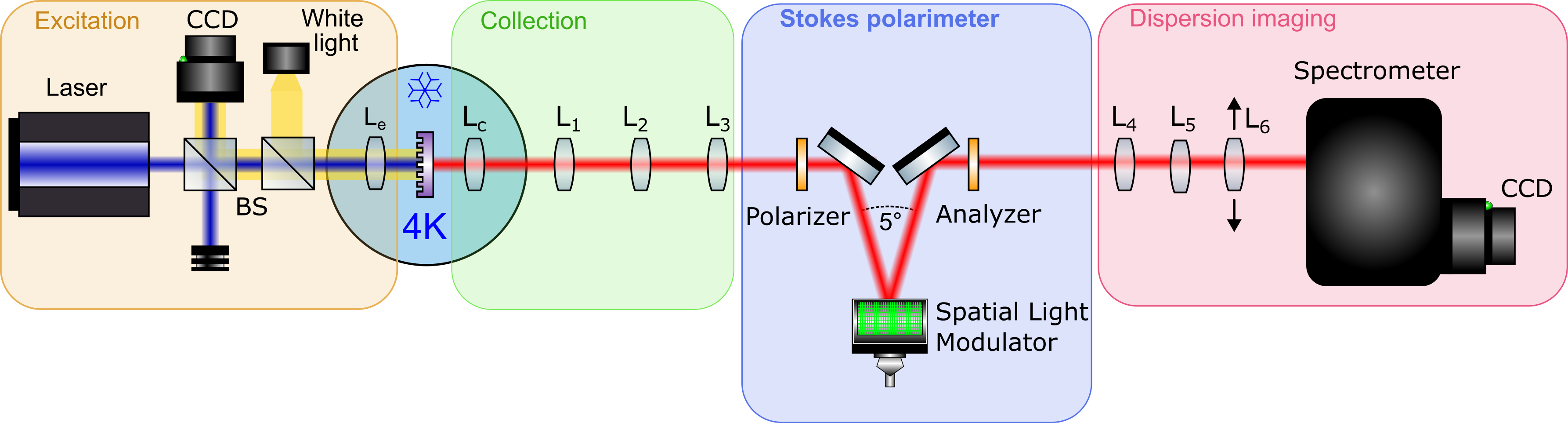}
    \caption{Detailed schematic of the experimental setup.}
    \label{fig:ExtDataSetup}
\end{figure*}

\begin{figure*}[p]
    \centering
    \includegraphics[width=1\textwidth]{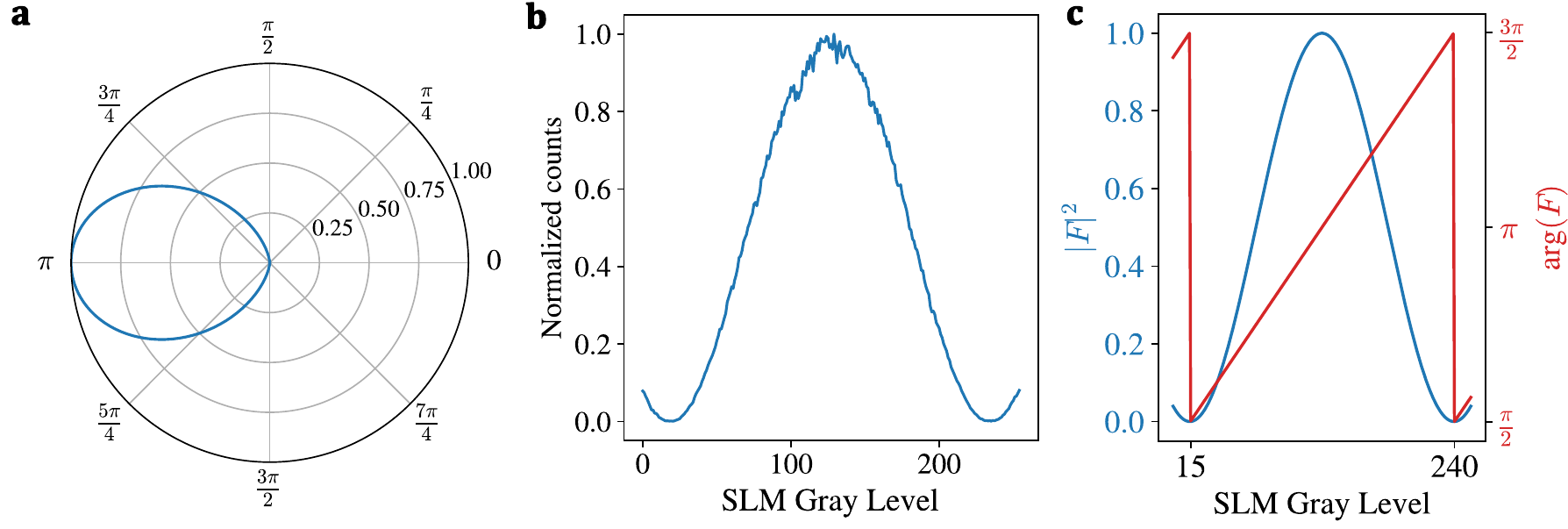}
    \caption{
    \textbf{a.}~Parametric plot showing $|F(\varphi)|^2$ as a function of $\text{arg}(F(\varphi))$.
    \textbf{b.}~Calibration of the polarimeter: transmitted intensity as a function of the uniform gray level on the SLM.
    \textbf{c.}~Calculated intensity ($|F|^2$) and phase ($\arg (F)$) response of the polarimeter as a function of the gray level on the SLM, from 0 to 255. 
    }
    \label{fig:ExtDataSLMcalib}
\end{figure*}

\begin{figure*}[p]
    \centering
    \includegraphics[width=\textwidth]{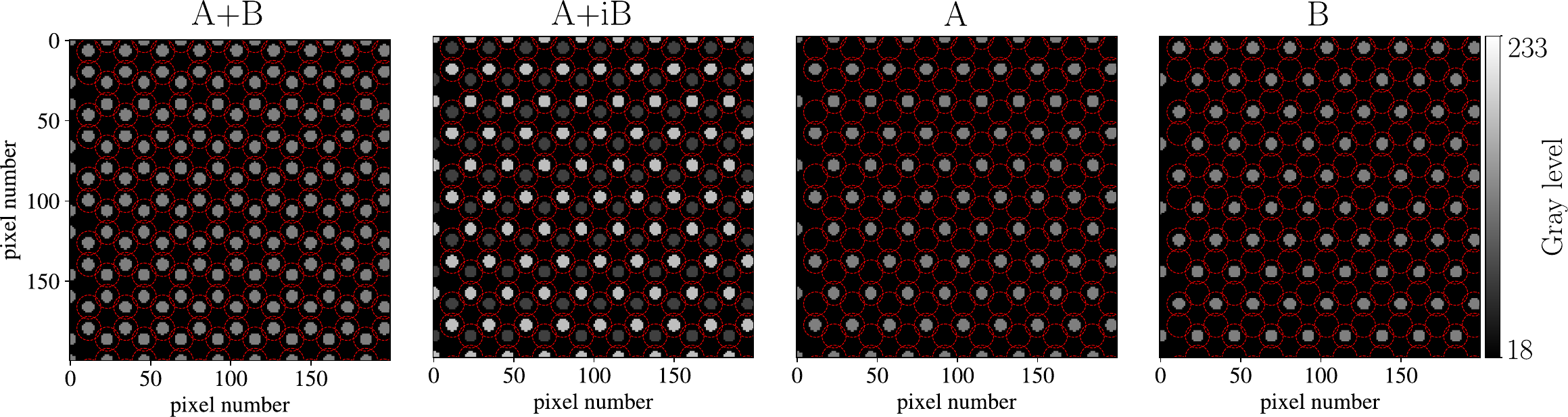}
    \caption{SLM mask patterns used for the four measurements of the Stokes polarimetry protocol. Dashed red circles indicate the targeted positions of the lattice sites.}
    \label{fig:ExtDataMasks}
\end{figure*}

\begin{figure*}[p]
    \centering
    \includegraphics[width=\textwidth]{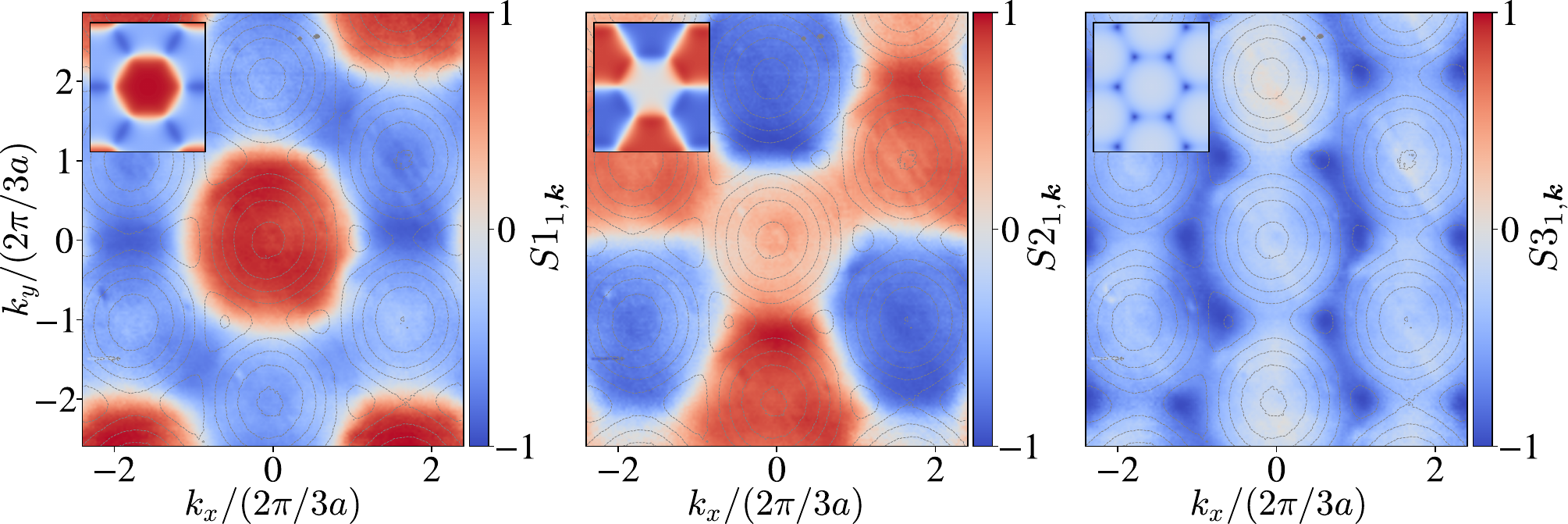}
    \caption{Reconstructed Stokes parameters $S1_{1,{\bm k}}$, $S2_{1,{\bm k}}$ and $S3_{1,{\bm k}}$ for the lower band of the staggered honeycomb lattice. Insets display the results of tight-binding simulations using the parameters $a = \SI{2.4}{\micro\meter}$ and $|\epsilon_0/t| = 1.26$.}
    \label{fig:ExtDataStokes}
\end{figure*}

\begin{figure*}[p]
    \centering
    \includegraphics[width=\textwidth]{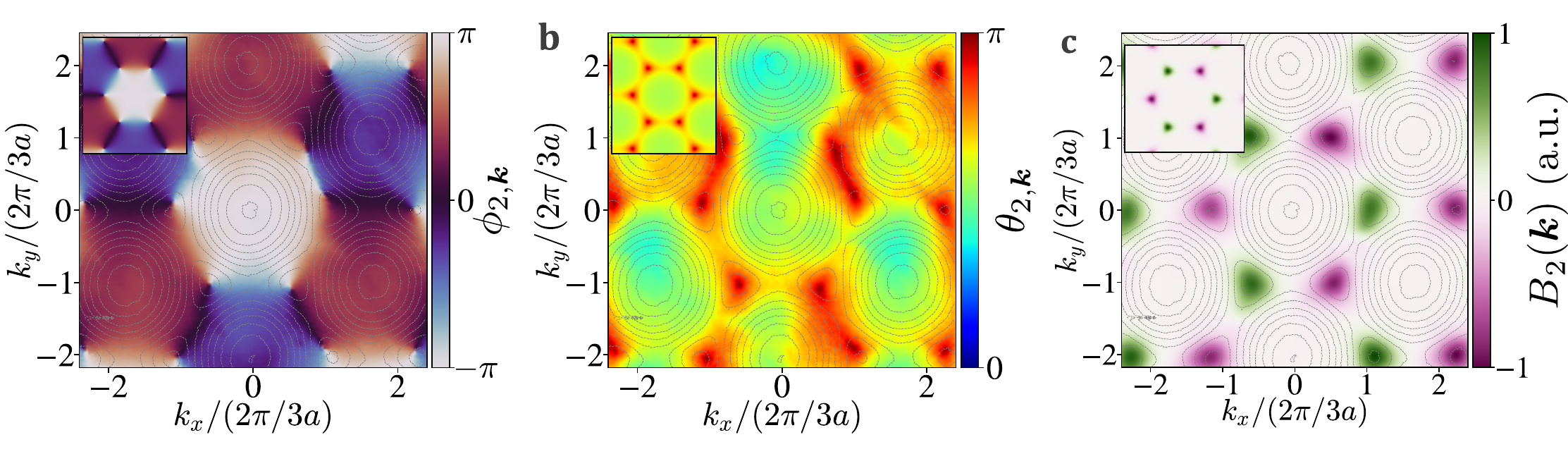}
    \caption{Reconstructed \textbf{a.}~azimuthal angle $\phi_{2,\bm{k}}$ and, \textbf{b.}~polar angle $\theta_{2,\bm{k}}$, and \textbf{c.}~resulting Berry curvature $B_2({\bm k})$ for the upper band of the staggered honeycomb lattice. Insets display the results of tight-binding simulations, using the parameters $a = \SI{2.4}{\micro\meter}$ and $|\epsilon_0/t| = 1.26$.}
    \label{fig:ExtDataEigenvectors}
\end{figure*}

\begin{figure*}[p]
    \centering
    \includegraphics[width=\textwidth]{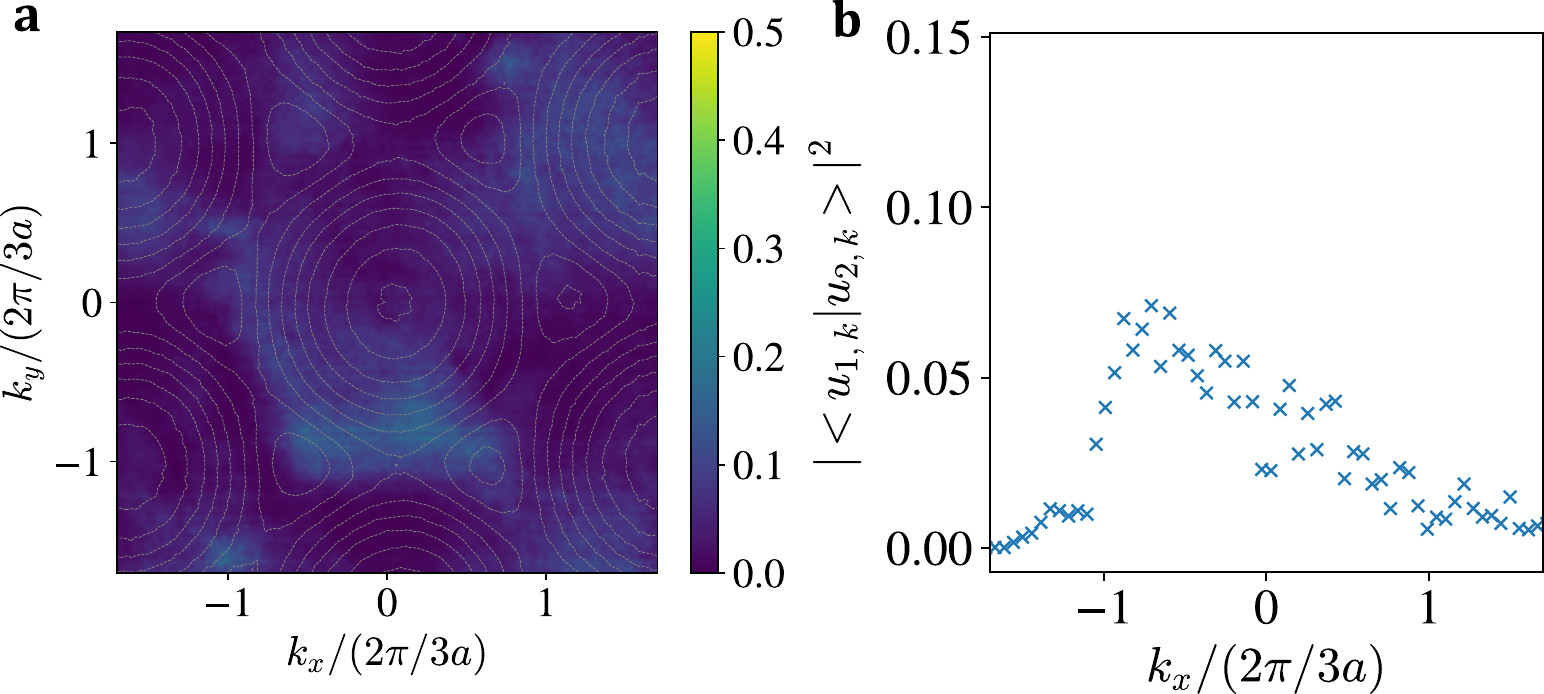}
    \caption{\textbf{a.}~Computed values of the scalar product $|\braket{u_{1,\bm{k}} | u_{2,\bm{k}}}|^2$ between the reconstructed eigenstates in the staggered honeycomb lattice. \textbf{b.}~Cut of the 2D graph shown in \textbf{a.} along $k_y = 0$.}
    \label{fig:ExtDataOrthogonality}
\end{figure*}

\clearpage

\begin{figure*}[p]
    \centering
    \includegraphics[width=\textwidth]{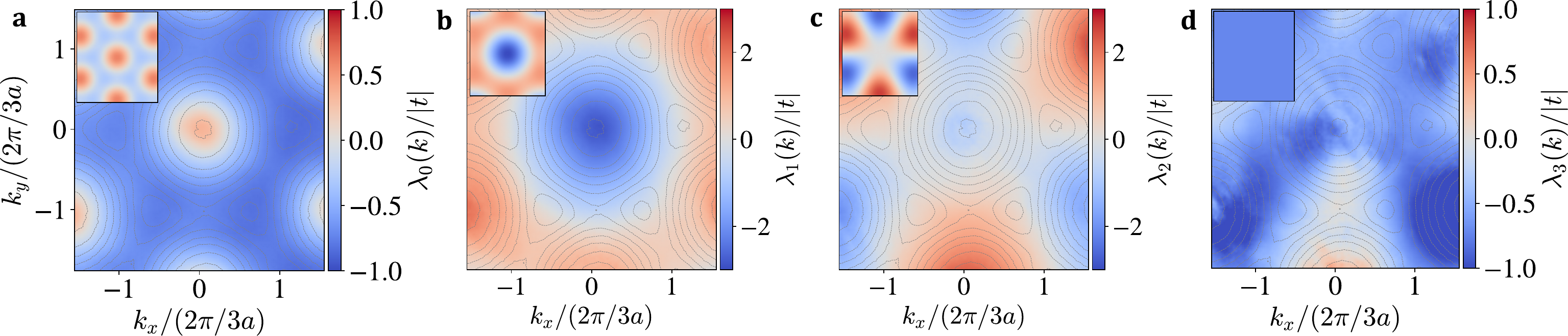}
    \caption{Reproducing the measured Hamiltonian by introducing a next-nearest-neighbor coupling $t'$ in the tight binding model. \textbf{a-d.}~Measured components $\lambda_i (k_x,k_y)$ of the Hamiltonian for the staggered honeycomb, with $i\in \{ 0,1,2,3\}$ (same data as shown in Fig.~\ref{fig:Fig5} of the main text). Insets show the results of tight-binding simulations, using the parameters $a = \SI{2.4}{\micro\meter}$, $|\epsilon_0/t| = 1.26$ and $|t'/t| = 0.1$.}
    \label{fig:ExtDataHamiltonian}
\end{figure*}

\begin{figure*}[p]
    \centering
    \includegraphics[width=\textwidth]{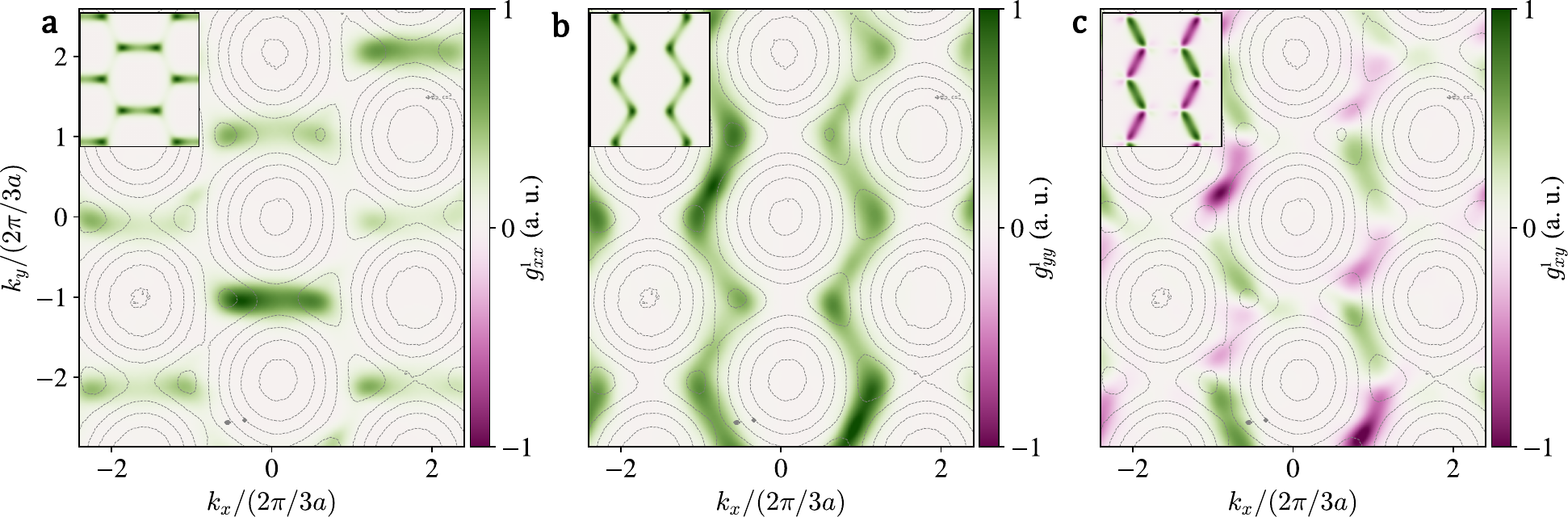}
    \caption{Measured components of the quantum metric $g_{ij}^1(k_x,k_y)$ of the lower band for the staggered honeycomb lattice. Insets display the results of tight-binding simulations, using the parameters $a = \SI{2.4}{\micro\meter}$ and $|\epsilon_0/t| = 1.26$.}
    \label{fig:ExtDataQGT}
\end{figure*}

\clearpage

\begin{figure*}[p]
    \centering
    \includegraphics[width=\textwidth]{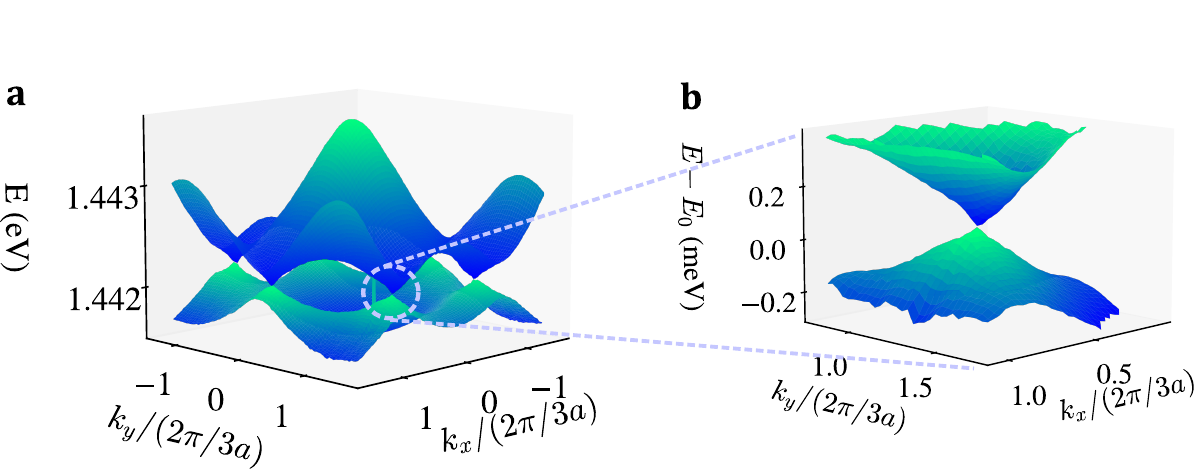}
    \caption{~\textbf{a.}~Three-dimensional representation of the measured band structure for the gapless honeycomb lattice, obtained by fitting the functions $\lambda_\pm(E)$ at each $\bm{k}$. \textbf{b.}~Zoom around a Dirac point of the measured band structure, demonstrating the ability of the sublattice Stokes polarimeter to resolve the band-touching point with sublinewidth precision.}
    \label{fig:ExtDataGraphene}
\end{figure*}

\clearpage



\renewcommand{\thefigure}{S\arabic{figure}}

\section*{Supplementary Information}

\textbf{The Stokes polarimeter presented in this work relies on controlling the interference between emissions from the $A$ and $B$ sublattices by applying a phase mask to the photoluminescence emission pattern. In the main text, we treat each lattice site as a point-like emitter with a well-defined phase, to which we apply uniform phase shifts. To do so, we discretize the emission pattern by selecting a small region within each micropillar using the SLM. In Section~A of this Supplementary Information, we detail the discretization procedure using the SLM mask, and the mapping of the 2D continuous lattice to a discrete tight-binding model. Section B analyzes the implications of this mapping on the measured Fourier-space emission. Finally, in Section~C, we study how the reconstruction of the Bloch eigenmodes is affected by the size of the selected regions and by potential misalignment between the SLM mask and the lattice image.}

\bigskip

To calculate the eigenmodes of the micropillar cavity, taking into account their two-dimensional continuous nature, we first decouple the longitudinal direction (orthogonal to the cavity mirrors) from the transverse plane, where the lattice geometry imposes translational symmetry. The photonic modes in the transverse plane are then obtained by solving the two-dimensional Schr\"odinger equation for the electromagnetic field amplitude $\Psi^{2D}_{n,\bm{k}} (\bm{r})$: 
\begin{equation*}
    \left[-\frac{\hbar^2}{2 m_c} \Delta + V(\bm{r}) \right] \Psi^{2D}_{n,\bm{k}} (\bm{r}) =  E_{n,\bm{k}} \Psi^{2D}_{n,\bm{k}} (\bm{r}) \, ,
\end{equation*}
where $m_c = \SI{3.0e-35}{\kilo\gram}$ is the effective mass of the cavity mode, extracted by fitting the parabolic dispersion of an unpatterned planar cavity. The potential $V(r)$ is a step-like honeycomb potential of amplitude $V_0 = \SI{600}{\milli\electronvolt}$, matching the exact geometry of the lithographic mask used to pattern the sample. This approach enables full simulations of the Stokes polarimeter, and allows us to quantitatively assess the validity of the eigenstate reconstruction method presented in the main text. 

\subsection{Mapping of the 2D lattice to a discrete tight binding model}
\label{Sup:SectionA}

Restricting the analysis to the two lowest photonic bands, in the spirit of linear combination of atomic orbital theory, we consider a basis of localized modes $\phi^{A}(\bm{r})$ and $\phi^{B}(\bm{r})$, each centered on a site of either the $A$ or the $B$ sublattice. Due to the lattice periodicity, the Bloch theorem applies, we can search eigenmodes of the continuous 2D system as follows:
\begin{equation*}
\Psi^{2D}_{n,\bm{k}} (\bm{r}) = 
\frac{1}{\sqrt{N_{\rm{tot}}}} \, e^{i \bm{k} \cdot \bm{r}} \sum_{j} \sum_{\sigma = A,B} u^{\sigma}_{n,\bm{k}} \,  \phi^{\sigma}(\bm{r} - \bm{r}_{j}^\sigma) \, ,
\end{equation*}
\noindent where $N_{\rm{tot}}$ is the total number of unit cells, $u^{\sigma}_{n,\bm{k}}$ are complex coefficients, and $\bm{r}_{j}^\sigma$ denotes the position of site $\sigma$ in unit cell $j$.

Our approach for measuring the eigenstates consists in using the SLM to select the emission of a small region of radius $R_m$ centered on each lattice site, and to apply a local amplitude attenuation or phase shift. The effect of the SLM mask on lattice site $\sigma$ in unit cell $j$ can be described by the function $m_{\sigma} (\bm r)$ defined as:
\begin{equation}
    m^\sigma(\bm{r}) = 
    c^\sigma \, \Theta \left( R_m - |\bm{r} - \bm{r}_j^\sigma | \right) \, ,
\end{equation}
\noindent where $\Theta(R)$ denotes the Heaviside step function, equal to 1 for positive arguments and 0 otherwise, and $c^{\sigma}$ is a complex coefficient encoding amplitude attenuation and phase shift applied to sublattice $\sigma$. This filtering operation effectively discretizes the emission pattern and enables site-resolved phase and amplitude control between sublattices, which is required for the operation of the polarimeter. Applying the mask to $\Psi^{2D}_{n,\bm{k}}$, we get the filtered 2D mode:
\begin{align*}
    \Psi_{n,\bm{k}} {\vert _ m} (\bm{r})
    & = \frac{1}{\sqrt{N_{\rm{tot}}}} \sum_{j}   \sum_{\sigma = A,B} c^\sigma \, u^{\sigma}_{n,\bm{k}} \, e^{i \bm{k} \cdot \bm{r}} \, \Theta \left( R_m - |\bm{r} - \bm{r}_j^\sigma | \right) \, \phi^{\sigma}(\bm{r} - \bm{r}_{j}^\sigma) \\
    & \simeq \frac{1}{\sqrt{N_{\rm{tot}}}} \sum_{j} \sum_{\sigma = A,B} c^\sigma \, u^{\sigma}_{n,\bm{k}} \, e^{i \bm{k} \cdot \bm{r}_j^\sigma} \, {\phi}^{\sigma}\vert_m(\bm{r} - \bm{r}_{j}^\sigma) \, ,
\end{align*}
\noindent where we have defined the truncated orbitals ${\phi}^{\sigma}\vert_m ({\bm r})$, obtained by multiplying the original orbital $\phi^{\sigma} ({\bm r})$ by the mask. In the second line of the equation, we have made the assumption that the Bloch phase factor varies slowly over the mask region, such that $e^{i \bm{k} \cdot \bm{r}} \approx e^{i \bm{k} \cdot \bm{r}_j^{\sigma}}$ within each mask. Under these conditions, we \sout{notice} obtain that the continuous system effectively reduces to a discrete model: in particular, when $c^{A} = c^{B} = 1$, $\Psi_{n,\bm{k}} (\bm{r})$ takes the form of the eigenfunctions of the $2 \times 2$ tight-binding Hamiltonian introduced in Eq.~1 of the main text. Therefore, this validates the mapping of the 2D continuous photonic lattice to the two-band tight-binding model.

\begin{figure}
    \centering
    \includegraphics[width=\textwidth]{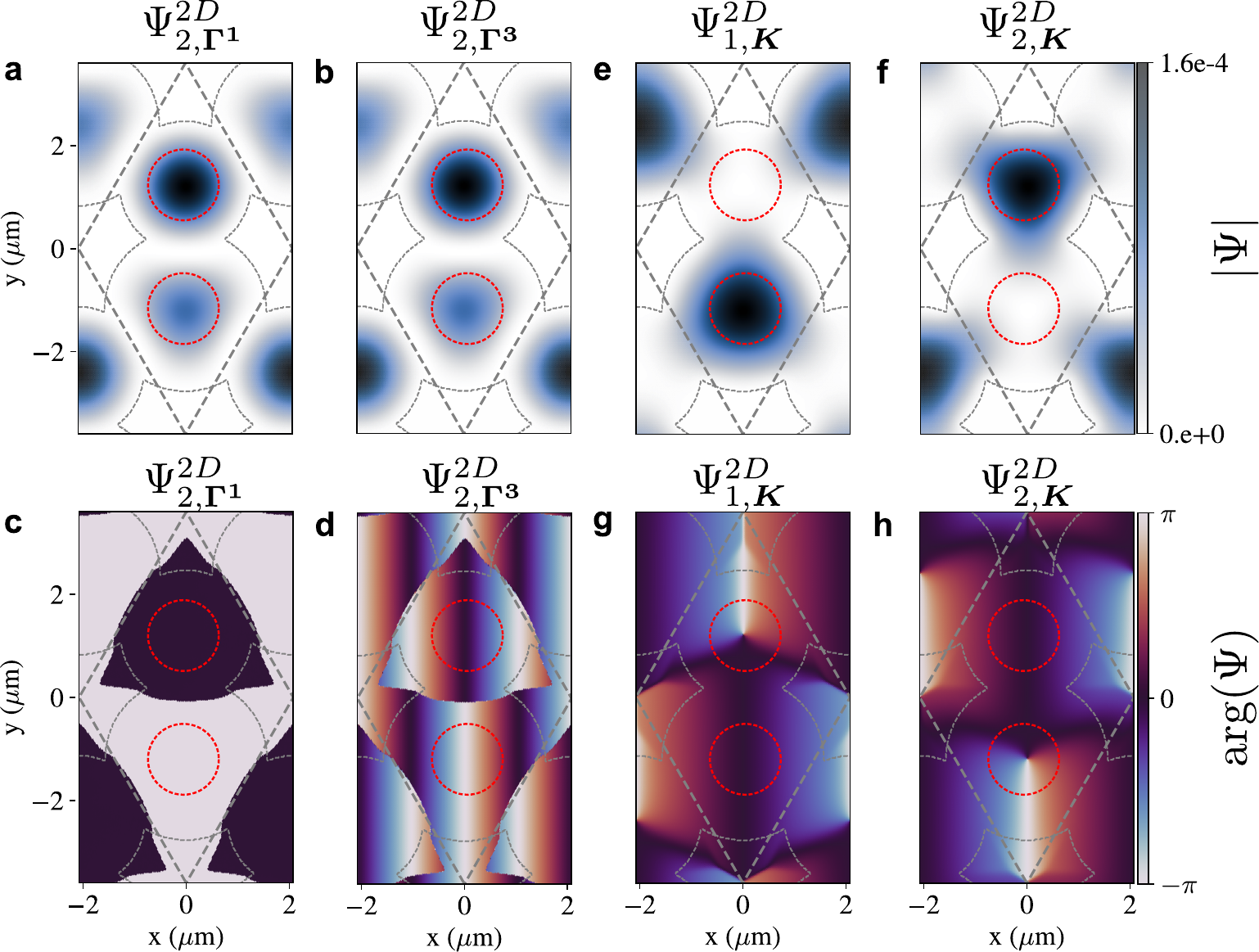}
    \caption{Bloch eigenfunctions of the staggered honeycomb lattice at various points in reciprocal space calculated by solving the two-dimensional Schr\"odinger equation. The top row shows the squared modulus of the modes, while the bottom row shows the corresponding phase profiles. Dashed red circles show the position and size of the SLM masks used in the experiment.
    \textbf{a,c}.~Bloch eigenfunction at the $\Gamma^1$ point of the first Brillouin zone for the upper band.
    \textbf{b,d}.~Bloch eigenfunction at the $\Gamma^3$ point of a third Brillouin zone ($k_y = 0$, and $k_x > 0$) for the upper band.
    \textbf{e,g}.~Bloch eigenfunction at a $K$-point in the first Brillouin zone for the lower band.
    \textbf{f,h}.~Bloch eigenfunction at a $K$-point in the first Brillouin zone for the upper band.}
    \label{Fig:Sup:Orbitals}
\end{figure}

To ensure that $e^{i \bm{k} \cdot \bm{r}}$ remains approximately constant within each region selected by the SLM mask, $|{\bm k}|$ and $R_m$ need to be taken sufficiently small so that $|{\bm k}| R_m \ll 1$. For example, for the mapping to hold across the first Brillouin zone, one must make sure that $R_m \ll 3a\sqrt{3}/(4\pi) \simeq \SI{1}{\micro\meter}$. In our experiments, we use $R_m = \SI{0.71}{\micro \meter}$, which ensures an accurate reconstruction of the eigenstates within the first Brillouin zone. This is illustrated in Fig.~\ref{Fig:Sup:Orbitals}a and Fig.~\ref{Fig:Sup:Orbitals}c, where we show the amplitude and phase of the two-dimensional wavefunction obtained by solving the two-dimensional Schr\"odinger equation for the upper band at the $\Gamma^1$ (the center of the first Brillouin zone). The anti-bonding character of the mode is clearly visible, with a characteristic $\pi$-phase shift between adjacent sites and vanishing amplitude at the junctions between sites. Importantly, the phase remains nearly uniform across the SLM mask area (see red dashed circle), thus enabling high-fidelity reconstruction of the Bloch eigenstates.

For larger values of ${\bm k}$, the phase can vary significantly across the area of a lattice site. This is illustrated in Fig.~\ref{Fig:Sup:Orbitals}b and Fig.~\ref{Fig:Sup:Orbitals}d, where we show the amplitude and phase of the two-dimensional wavefunction for the upper band at the $\Gamma^3$ point (the center of the third Brillouin zone). While the anti-bonding nature of the mode is still clearly observed, intra-site phase oscillations become apparent. The use of the SLM mask, centered on each site, enables us to partially suppress these oscillations by filtering out some of the phase spatial variations. However, residual variations of the phase within the mask area may lead to deviations from the ideal tight-binding approximation, especially at higher-order Brillouin zones. These effects are further discussed in Section~\ref{Sup:SectionC}, where we present simulations illustrating the impact of the mask size and alignment on the accuracy of the eigenmode reconstruction.

\subsection{Fourier space emission}
\label{Sup:SectionB}

Following the filtering of the eigenstates by the SLM, we now analyze the resulting Fourier-space emission. The Fourier transform of $\Psi_{n,\bm{k}} {\vert _ m} (\bm{r})$ yields:
\begin{align*}
    \varphi_{n,{\bm k}}({\bm q})
    & = \frac{1}{\sqrt{N_{\rm tot}}} \sum_j \sum_{\sigma = A,B} c^\sigma \, u_{n,{\bm k}}^\sigma \, e^{i {\bm k} \cdot {\bm r}_j^{\sigma}} \iint e^{ - i {\bm q} \cdot {\bm r}} \, {\phi}^{\sigma}\vert_m(\bm{r} - \bm{r}_{j}^\sigma) \, d{\bm r} \\
    & = \frac{1}{\sqrt{N_{\rm tot}}} \sum_{\sigma = A,B} c^\sigma \, u_{n,{\bm k}}^\sigma \, \left( \sum_j e^{i ({\bm k}-{\bm q}) \cdot {\bm r}_j^{\sigma}} \right) \iint_{|{\bm r}| < R_m} e^{ - i {\bm q} \cdot {\bm r}} \, {\phi}^{\sigma}(\bm{r}) \, d{\bm r} \\
    & = \frac{1}{\sqrt{N_{\rm tot}}} \sum_{\sigma = A,B} c^\sigma \, u_{n,{\bm k}}^\sigma \, S^{\sigma}({\bm k} - {\bm q}) \, {\tilde{\phi}}^{\sigma}\vert_m(\bm{q}) \, ,
\end{align*}
\noindent where $S^{\sigma}({\bm q}) = \sum_j e^{i {\bm q} \cdot {\bm r}_j^\sigma}$ is a structure factor, and ${\tilde{\phi}}^{\sigma}\vert_m(\bm{q})$ is the Fourier transform of $\phi^{\sigma}({\bm r})$ truncated to $|{\bm r}| < R_m$. For sufficiently large $N_{\rm tot}$, $S^{\sigma}({\bm k} - {\bm q})$ is sharply peaked around ${\bm k} = {\bm q}$, leading to a highly directional emission along the ${\bm k}$ direction of amplitude:
\begin{align*}
    \varphi_{n,{\bm k}}({\bm k})
    & = \sqrt{N_{\rm tot}} \sum_{\sigma = A,B} c^{\sigma} \,  u^{\sigma}_{n,\bm{k}} \, {\tilde{\phi}}^{\sigma}\vert_m(\bm{k}) \, .
\end{align*}

In the case of the graphene lattice, where all sites are identical, we have $\phi^{A}(\bm{r}) = \phi^{B}(\bm{r}) = \phi^{0}(\bm{r})$. Therefore:
\begin{equation*}
    \varphi_{n,{\bm k}} = \sqrt{N_{\rm tot}} \, {\tilde{\phi}}^0\vert_m(\bm{k}) \sum_{\sigma = A,B} c^{\sigma} \, u^{\sigma}_{n,\bm{k}} \, .
\end{equation*}
\noindent The field amplitude generated by the mode is then:
\begin{equation*}
    \mathcal{\tilde{E}}_{n,\bm{k}} (E) \propto c^{A} \, u_{n,{\bm k}}^{A} + c^B \, u_{n,{\bm k}}^{B} \, .
\end{equation*}
In particular, in the configuration where we use the SLM to select the center of each lattice sites without applying any phase shift nor attenuation ($c^A = c^B = 1$), we recover the expression used in the main text:
\begin{equation*}
    \mathcal{\tilde{E}}_{n,\bm{k}} (E) = \mathcal{A}_{n,\bm{k}} (E) \, (u_{n,{\bm k}}^{(A)} +u_{n,{\bm k}}^{(B)}) \, e ^{-i E t / \hbar} \, .
\end{equation*}

For the staggered honeycomb lattice, we verify numerically that the condition ${\tilde{\phi}}^{A}\vert_m(\bm{k}) \simeq {\tilde{\phi}}^{B}\vert_m(\bm{k})$ remains valid. To do so, we compute the localized orbitals $\phi^{A} (\bm{r})$ and $\phi^{B} (\bm{r})$ using the fact that, at the $K$ point, the Bloch eigenfunctions are fully localized on a single sublattice:
\begin{align*}
\Psi^{2D}_{1,\bm{K}} (\bm{r})
& = \frac{1}{\sqrt{N_{\rm{tot}}}} \, e^{i \bm{K} \cdot \bm{r}} \sum_{j} u^{A}_{n,\bm{K}} \,  \phi^{A}(\bm{r} - \bm{r}_{j}^A) \, , \\
\Psi^{2D}_{2,\bm{K}} (\bm{r})
&=\frac{1}{\sqrt{N_{\rm{tot}}}} \, e^{i \bm{K} \cdot \bm{r}} \sum_{j} u^{B}_{n,\bm{K}} \,  \phi^{B}(\bm{r} - \bm{r}_{j}^B) \, .
\end{align*}
\noindent We solve the 2D Schr\"odinger equation with $R_A = \SI{1.57}{\micro\meter}$ and $R_B = \SI{1.27}{\micro\meter}$ and $a=\SI{2.4}{\micro\meter}$. The resulting orbitals are shown in Fig.~\ref{Fig:Sup:Orbitals}(e-h), where we clearly observe the complete localization of the wavefunctions on the $A$ or $B$ sublattice. Interestingly, the vanishing of the wavefunction amplitude $| \Psi |$ on one sublattice observed in Fig.~\ref{Fig:Sup:Orbitals}(e-f) gives rise to vortices in the phase profiles $\arg ( \Psi ) $, as shown in Fig.~\ref{Fig:Sup:Orbitals}(g-h).

We then compute the integrals of $\phi^{\sigma} (\bm{r})$ truncated to the SLM mask area, and find:
\begin{equation*}
    \frac{ \left|\iint_{|{\bm r}|< R_m} \phi^{A}(\bm{r}) \, d\bm{r} \right| -  \left|\iint_{|{\bm r}|< R_m} \phi^{B}(\bm{r}) \, d\bm{r} \right|}{ \left|\iint_{|{\bm r}|< R_m} \phi^{A}(\bm{r}) \, d\bm{r} \right| +  \left|\iint_{|{\bm r}|< R_m} \phi^{B}(\bm{r}) \, d\bm{r} \right|} = 0.013 \, .
\end{equation*}
\noindent This confirms that the difference between the two orbitals is negligible, and the assumption ${\tilde{\phi}}^A\vert_m (\bm{k}) \simeq {\tilde{\phi}}^B\vert_m (\bm{k})$ is therefore valid.

\begin{figure}
    \centering
    \includegraphics[width=\textwidth]{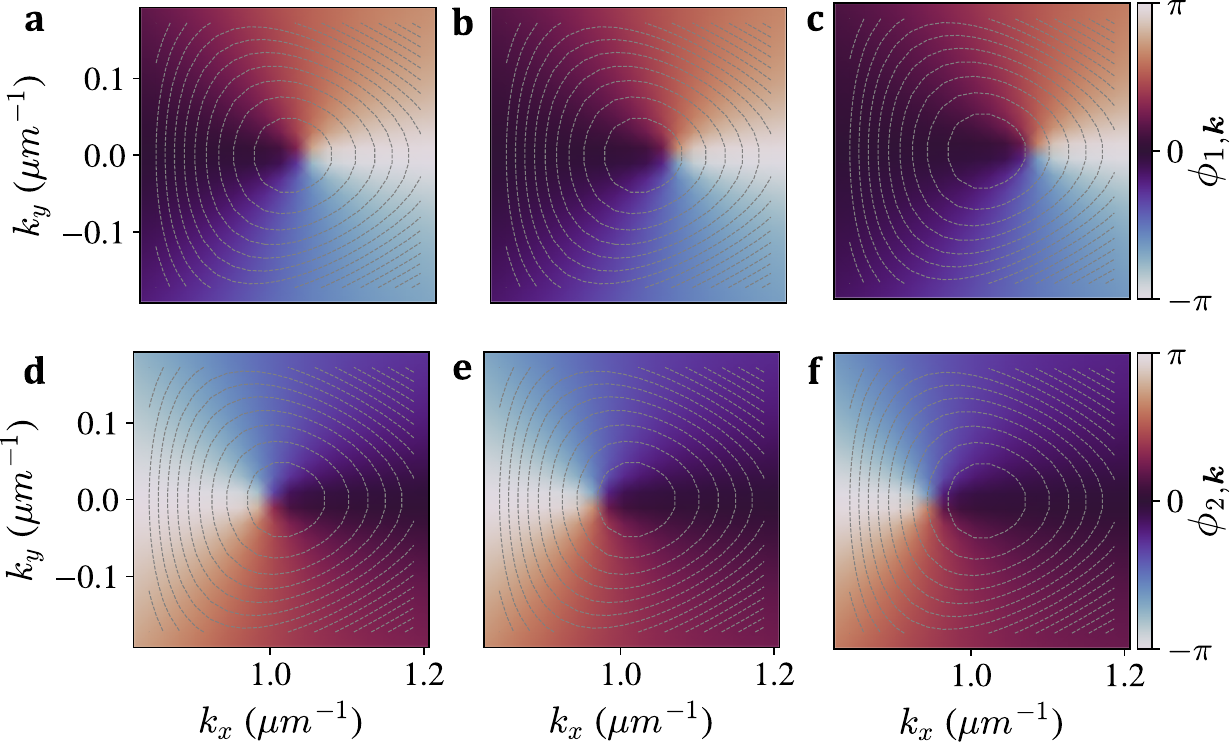}
    \caption{Calculated values of the azimuthal angle $\phi_{1, \bm{k}}$ (upper line) and $\phi_{2,\bm{k}}$ (lower line) obtained by solving the two-dimensional Schr\"odinger equation, for the lower and upper band using three values of the mask radius $R_m$. (\textbf{a,d})  $R_m = \SI{0.715}{\micro\meter}$, (\textbf{b,e}) $R_m = \SI{1.14}{\micro\meter}$ and (\textbf{c,f}) t $R_m = \SI{1.43}{\micro\meter}$. The grey dashed lines trace the bands isoenergy lines.}
    \label{Fig:Sup:MaskSize}
\end{figure}

\begin{figure}
    \centering
    \centering
    \includegraphics[width=\textwidth]{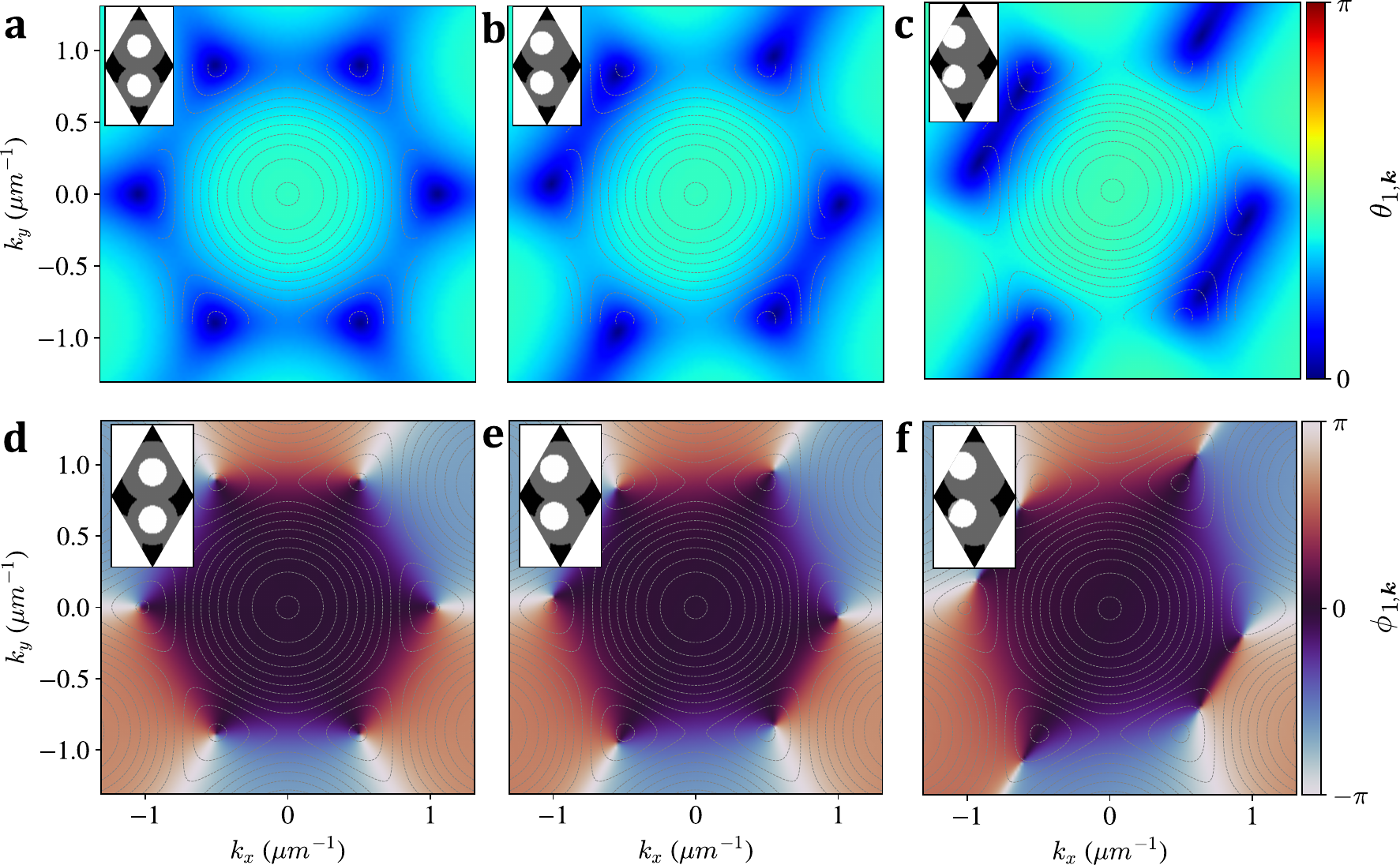}
    \caption{Values of the polar and azimuthal angles $\theta_{1,\bm{k}}$ (upper line) and $\phi_{1,\bm{k}}$ (lower line) calculated for the lower band by solving the two-dimensional Schr\"odinger equation. We consider three configurations of the mask alignment: (a-d) Perfect mask alignment, (b-e) the mask is displaced along the inter-poillar segment by $15\%$ of the center to center distance, (c,f) the mask is displaced along the inter-pillar segment by $30\%$ of the center to center distance. Insets represent in white the mask position and in grey the unit cell. The grey dashed lines trace the bands isoenergy lines.}
    \label{Fig:Sup:Misalignment}
\end{figure}

\subsection{Influence of the mask area and its misalignment on the eigenmode reconstruction}
\label{Sup:SectionC}

In this Section, we present simulations of the Stokes polarimetry experiment for different implementations of the SLM mask, and evaluate the accuracy of the Bloch eigenstates reconstruction. To do so, we apply the four SLM masks of the protocol to the computed eigenmodes of the two-dimensionnal Schr\"odinger equation.

We first analyse the effect of the mask radius $R_m$, defining the size of the area selected by the SLM on each lattice site. Fig.~\ref{Fig:Sup:MaskSize} shows the calculated values of the azimuthal angle for the lower and upper band using three different values of $R_m$. As $R_m$ increases, we observe that the reconstructed phase vortices shift away from the Dirac points, in opposite direction for each band. This deviation from the tight-binding prediction originates from the growing variations of the phase factor $e^{i \bm{k} \cdot \bm{r}}$ within the mask area (see Section~\ref{Sup:SectionA}). For the value $R_m = \SI{0.715}{\micro\meter}$, which closely matches the value used in the experiments, the reconstructed vortices remain centered near the Dirac points, confirming the validity of the mapping to the discrete tight-binding model. 

Finally, in Fig.~\ref{Fig:Sup:Misalignment}, we study the impact of a misalignment between the SLM mask and the lattice image. We compare the computed values of the polar and azimuthal angles $\theta_{1,\bm{k}}$ and $\phi_{1,\bm{k}}$ reconstructed for the lower band under different alignment conditions. As the mask becomes increasingly misaligned, we observe both a displacement of the phase vortices and an overall distortion the Brillouin zone shape. This highlights the importance of the precise alignment procedure described in the Method section, which is essential for the accurate implementation of the Stokes polarimeter.

In summary, provided that the SLM mask is sufficiently small and properly aligned, the continuous 2D photonic lattice can be reliably mapped onto a discrete tight-binding model. This enables the high-fidelity reconstruction of the Bloch eigenstates as shown in the main text.

\end{document}